\documentclass[useAMS,usegraphicx,usenatbib]{mn2e}
\usepackage{graphicx}
\usepackage{amsmath}
\usepackage{amssymb}
\usepackage{subfigure}
\usepackage{url}
\usepackage{multirow}
\usepackage[linkcolor=blue,colorlinks=true,breaklinks,citecolor=blue,urlcolor=blue]{hyperref} 
\newcommand{\km}{\rm\thinspace km}

\newcommand{\cm}{\rm\thinspace cm}

\newcommand{\s}{\rm\thinspace s}
\newcommand{\ks}{\rm\thinspace ks}

\newcommand{\Msun}{\hbox{$\rm\thinspace M_{\odot}$}}

\newcommand{\keV}{\rm\thinspace keV}
\newcommand{\eV}{\rm\thinspace eV}
\newcommand{\erg}{\rm\thinspace erg}

\newcommand{\ergpcmsqps}{\hbox{$\erg\cm^{-2}\s^{-1}\,$}}

\newcommand{\cts}{\rm\thinspace ct}
\newcommand{\ctsps}{\hbox{$\cts\s^{-1}\,$}}
\newcommand{\ergcmps}{\hbox{$\erg\cm\ps\,$}}
\newcommand{\kmps}{\hbox{$\km\s^{-1}\,$}}

\newcommand{\psqcm}{\hbox{$\cm^{-2}\,$}}

\newcommand{\ps}{\hbox{$\s^{-1}\,$}}

\newcommand{\rg}{\rm\thinspace $r_\mathrm{g}$}

\title[The evolving corona of Mrk~335]{Driving extreme variability: The evolving corona and evidence for jet launching in Markarian 335}
\author[D. R. Wilkins \& L.C. Gallo]{D. R. Wilkins\thanks{E-mail: drw@ap.smu.ca}\thanks{CITA National Fellow} and L. C. Gallo\\Department of Astronomy \& Physics, Saint Mary's University, Halifax, NS. B3H 3C3 Canada}
\begin{document}

\date{Accepted 2015 January 22.  Received 2015 January 21; in original form 2014 September 23}

\pagerange{\pageref{firstpage}--\pageref{lastpage}} \pubyear{2014}

\maketitle

\label{firstpage}

\begin{abstract}
Variations in the X-ray emission from the narrow line Seyfert 1 galaxy, Markarian 335 (Mrk~335), are studied on both long and short timescales through observations made between 2006 and 2013 with \textit{XMM-Newton}, \textit{Suzaku} and \textit{NuSTAR}. Changes in the geometry and energetics of the corona that give rise to this variability are inferred through measurements of the relativistically blurred reflection seen from the accretion disc. On long timescales, we find that during the high flux epochs the corona has expanded, covering the inner regions of the accretion disc out to a radius of $26_{-7}^{+10}$\rg. The corona contracts to within 12\rg\ and 5\rg\ in the intermediate and low flux epochs, respectively. While the earlier high flux observation made in 2006 is consistent with a corona extending over the inner part of the accretion disc, a later high flux observation that year revealed that the X-ray source had become collimated into a vertically-extended jet-like corona and suggested relativistic motion of material upward. On short timescales, we find that an X-ray flare during a low flux epoch in 2013 corresponded to a reconfiguration from a slightly extended corona to one much more compact, within just $2\sim 3$\rg\ of the black hole. There is evidence that during the flare itself, the spectrum softened and the corona became collimated and slightly extended vertically as if a jet-launching event was aborted. Understanding the evolution of the X-ray emitting corona may reveal the underlying mechanism by which the luminous X-ray sources in AGN are powered.
\end{abstract}


\section{Introduction}
Markarian 335 (Mrk~335) is a particularly interesting example of a narrow line Seyfert 1 (NLS1) galaxy, harbouring a supermassive black hole of mass $2.6\times 10^7$\Msun\ \citep{grier+12}, whose X-ray emission has been studied on numerous occasions by missions as far back as \textit{UHURU} and, more recently, \textit{XMM-Newton}, \textit{Suzaku} and \textit{NuSTAR}. Each observation of Mrk~335 found new phenomena, from relativistically blurred reflection \citep{gallo+13} to changing intrinsic absorption from outflowing winds \citep{longinotti+13}.

Over the last 15 years, the flux from Mrk~335 has varied by more than an order of magnitude, with early observations from 2000 to 2006 seeing a bright X-ray source in a high flux state. The flux was then found to have dropped by a factor of 10 when Mrk~335 was observed by \textit{XMM-Newton} in 2007 before recovering into an intermediate flux state by 2009 during which the source was found to be transitioning from lower to higher luminosity. Most recently, monitoring of Mrk~335 by the \textit{Swift} satellite found the flux to have once again dropped to a similar level to that seen in 2009, hence target of opportunity (ToO) observations were triggered using \textit{NuSTAR} \citep{parker_mrk335} and, simultaneously with part of this, \textit{Suzaku} \citep{gallo+14}.

The X-ray spectra of Mrk~335, ranging from the early high-flux observations with \textit{XMM-Newton} \citep{crummy+06}, \textit{Suzaku} \citep{larsson+08} and even \textit{ASCA} \citep{ballantyne+01} to the low \citep{grupe+08} and intermediate flux observations \citep{gallo+13} from later years can be explained by X-ray continuum emission from a corona of energetic particles surrounding the central black hole that illuminates the accretion disc of material spiralling inward \citep{george_fabian}. This leads to X-ray reflection by the processes of Compton scattering, photoelectric absorption and the emission of fluorescence lines and bremsstrahlung \citep{ross_fabian}. The reflection spectrum, including the prominent 6.4\keV\ K$\alpha$ emission line of iron, is blurred by Doppler shifts and relativistic beaming due to the orbital motion of material in the accretion disc as well as by gravitational redshifts in the strong gravitational field around the black hole \citep{fabian+89,laor-91}.

In addition to the blurred reflection from the accretion disc, absorption features are seen in the X-ray spectrum that are attributed to material outflowing from the central regions of the AGN. \citet{longinotti+13} find significant spectral features due to absorption by ionised outflowing material, attributed to a wind launched from the surface of the accretion disc during the 2009 intermediate flux epoch and that some of this absorption remains during the 2006 and 2007 high and low flux epochs observed by \textit{XMM-Newton}. On the other hand, \citet{larsson+08} find that no intrinsic absorption is required to model the high flux spectrum recorded by \textit{Suzaku} later in 2006.

The soft (0.5-10\keV) X-ray spectrum of Mrk~335 can be equally well explained by the partial covering of a primary X-ray source by absorbing material along the line of sight, with no detection of relativistically blurred reflection from the accretion disc \citep{longinotti+07,oneill+07,grupe+08}. \citet{gallo+14}, however, find that in order to self-consistently describe both high- and low-flux observations of Mrk~335 made with \textit{Suzaku} by changes in a partially-covering absorber is challenging. Extreme parameters are required, with almost complete covering by Compton-thick material being required in the low flux state with little to no absorption in the high flux state. Variation in the primary X-ray continuum that gives rise to relativistically blurred reflection from the accretion disc provides a much more natural description of the data.

Compelling evidence for relativistically blurred reflection from the accretion disc in Mrk~335 comes from the detection of X-ray reverberation time lags between the variability in the X-ray continuum and the corresponding variations in the reflected X-rays \citep{kara+13}. The measured time lag corresponds to the light crossing time between the X-ray emitting corona and the inner regions of the accretion disc, while the variation in time lag as a function of X-ray energy is consistent with that seen in many NLS1 galaxies \citep{demarco+2012,reverb_review} and is indicative of the most redshifted emission in the wing of the 6.4\keV\ iron K$\alpha$ line emanating from the inner regions of the accretion disc where gravitational redshift is most extreme, closer to the primary X-ray source, while a longer lag is seen from the core of the line, consisting of X-rays reflected from the outer disc. At the same time, \citet{parker_mrk335} clearly detect the Compton hump characteristic of X-ray reflection in broadband spectra obtained by \textit{NuSTAR} finding the spectrum to be well described by the relativistically blurred reflection of X-rays originating from a compact corona within just a few gravitational radii of the black hole.

Recent detailed analysis of the X-rays reflected from the accretion disc has enabled the measurement of the geometry of the corona that is illuminating the disc, particularly through the emissivity profile of the accretion disc \citep{1h0707_emis_paper,understanding_emis_paper} and the spectrum of reverberation time lags as a function of both the frequency of variability and X-ray energy \citep{lag_spectra_paper,cackett_ngc4151}. Much remains unknown, however, about the physical processes occurring within the coron\ae\ in AGN. Measuring the evolution in the corona as the X-ray luminosity varies on both long and short timescales will give important insight to the structure and physical processes within the X-ray emitting region. Understanding the variations in the corona may also reveal the mechanism by which energy is liberated from the accretion flow and injected into the corona to accelerate the particles in order to power some of the most luminous objects we see in the Universe. Given its great variability over orders of magnitude in luminosity, the variable nature of outflows and other phenomena found in this source, and it being one of the closer NLS1 galaxies meaning that even in during the low flux epochs, a respectable count rate is measured, Mrk~335 is the ideal source in which to study the variability of the X-ray emitting corona to learn about the environment and processes therein.

In this paper, we analyse the corona of Mrk~335 over high and low flux epochs through X-ray spectra observed between 2006 and 2013. We begin by outlining the means by which measurements can be made of the corona from X-ray observations before studying variations inthe corona through the reflection of X-rays from the accretion disc over both long timescales and short timescales through the 2013 low flux observation made with \textit{Suzaku}.

\section{Observations and Data Reduction}
In order to probe the evolution of the corona that may lead to the extreme variability exhibited by Markarian 335, data from five sets of observations with the X-ray observatories \textit{XMM-Newton} \citep{xmm} and \textit{Suzaku} \citep{suzaku} during which both the observed spectrum and flux varies greatly, shown in Table~\ref{obs.tab} were analysed.

\begin{figure}
\centering
\includegraphics[width=85mm]{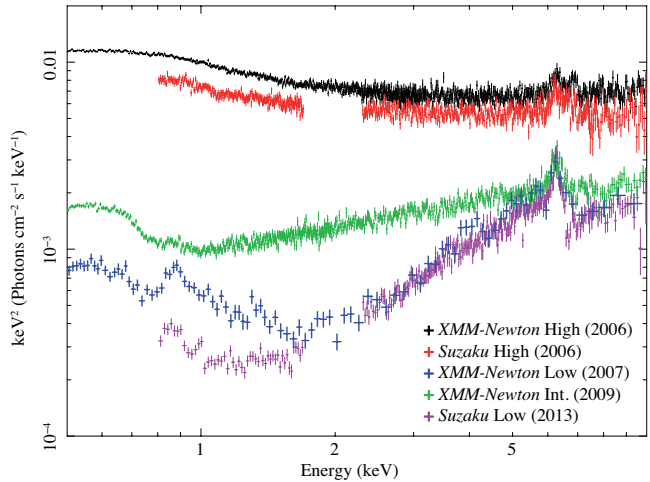}
\caption[]{Unfolded spectra of Markarian 335 during the five epochs, observed with \textit{XMM-Newton} and \textit{Suzaku} between 2006 and 2013. The instrument response is unfolded from the recorded spectra using \textsc{xspec}, assuming a featureless, constant model spectrum across all energies (\textit{i.e.} a power law with index zero), to enable the spectra and flux levels to be quickly compared across the epochs, even through they were recorded with different instruments.}
\label{unfolded_spectra.fig}
\end{figure}

\textit{XMM-Newton} observed Mrk~335 during high, low and intermediate flux epochs in 2006, 2007 and 2009, respectively. We do not use the high flux observation of Mrk~335 obtained by \textit{XMM-Newton} in 2000 as analysis suggests that this shorter exposure, during which the EPIC pn camera was operated in full frame mode, suffers from a significant degree of pile-up. \textit{Suzaku} has observed Mrk~335 on two occasions; a high flux epoch during 2007 and a target of opportunity (ToO) observation during 2013, triggered from \textit{Swift} monitoring to observe Mrk~335 at a low flux level, simultaneously with \textit{NuSTAR} \citep{nustar}. The 2013 \textit{Suzaku} observation shows similar characteristics to the 2007 low flux epoch observed by \textit{XMM-Newton}, but provides a considerable improvement in the quality of the data, offering around eight times as many photon counts from the source. The spectra during each of these observations, with the instrument response unfolded, are shown in Fig.~\ref{unfolded_spectra.fig} and allow a comparison to be made between the different epochs.

\subsection{XMM-Newton}
We use the spectra recorded by the just the EPIC pn camera on board \textit{XMM-Newton} \citep{xmm_strueder}, due to its enhanced effective area over the EPIC MOS cameras to allow for finer energy binning in the recorded spectra. Since Mrk~335 is a bright source, the EPIC pn camera provides sufficient counts for detailed analysis of the X-rays reflected from the accretion disc, while meaning we need not be concerned by the cross-calibration of the instruments, particularly when decomposing the relativistically blurred reflection from the accretion disc to measure the emissivity profile.

Data were reduced using the \textit{XMM-Newton} \textsc{Science Analysis System} (\textsc{sas}) v13.5.0 using the most recent calibration data for the observations in question available at the time of writing. After initial reduction of the event lists and removal of background flares, the source spectra were extracted from a circular region of the detector, centred on the point source, 35\,arcsec in diameter. Corresponding background spectra were extracted from a region of the same size, on the same chip as the source. The spectra were binned using the \textsc{grppha} tool such that there were at least 25 counts in each spectral bin and that the errors are approximately Gaussian. The photon redistribution matrices (RMF) and ancillary response matrices (ARF), encoding the effective area as a function of energy, were computed using the \textsc{sas} tasks \textsc{rmfgen} and \textsc{arfgen}, following the standard procedure.

The \textit{XMM-Newton} observation of the 2009 intermediate flux epoch was divided into two orbits between which differences were observed in the X-ray spectrum \citep{gallo+13}. During the first orbit, the count rate rose from $2.3\pm0.4$\ctsps\ to $4.4\pm0.6$\ctsps\ and remained high during the second orbit, averaging $4\pm1$\ctsps. As well as considering the two orbits separately, the combined spectrum, summing the source and background counts from the two orbits (using averaged response matrices) was considered in order to to maximise the number of photon counts for detailed analysis of the X-ray reflection spectrum and the accretion disc emissivity profile, measuring the average properties of the source over the two orbits.

\subsection{Suzaku}
Mrk~335 was observed by \textit{Suzaku} in the HXD-nominal position in 2006 and in the XIS-nominal position in 2013. Spectra were extracted from observations by the \textit{X-ray imaging spectrometer} (XIS) CCDs on board \textit{Suzaku}. Unscreened event lists were reprocessed and filtered using the latest calibration data available with \textsc{aepipeline}, then spectra were extracted from a circular area, 250\,arcsec in diameter, centred on the point source. Corresponding background spectra were extracted from regions of the same size on each detector. The photon redistribution matrices (RMF) were generated for the observations using \textsc{xisrmfgen} and ancillary response matrices (ARF) encoding the effective area were calculated using the ray tracing tool \textsc{xissimarfgen}.

After checking for consistency, the spectra from the front-illuminated CCDs, XIS0 and XIS3 (and XIS2 during the 2006 obsveration, which was still functioning at this time), were combined into a single spectrum using the tool \textsc{addascaspec}. Due to uncertainties in the calibration, energy channels below 0.8\keV\ and between 1.7 and 2.3\keV\ were excluded from spectral fitting. During the 2013 low flux observation, strong background emission from Ni K$\alpha$ is detected relative to the low source flux, therefore energy channels between 7.4 and 7.8\keV\ were also excluded from the 2013 observation. XIS spectra were fit over the energy band $0.8-12.0$\keV.

The spectra were binned using \textsc{grppha} such that there were at least 25 X-ray counts in each spectral bin. We do not use the data from the back-illuminated XIS1 detector in order to avoid systematic errors arising from the cross-calibration of the front- and back-illuminated CCDs when conducting detailed analysis of the X-ray reflection spectra and the accretion disc emissivity profiles, choosing to use the combined spectra from the front-illuminated CCDs to maximise the number of available counts.

In order to extend coverage to higher energy, spectra were also extracted from the PIN component of the Hard X-ray Detector (HXD). PIN spectra were extracted from the HXD data following reprocessing and filtering of the unscreened event lists, using the latest calibration data available. The non X-ray background (NXB) computed from the simulated event list obtained from the calibration database (\textsc{caldb}) was compared to the background spectrum measured during periods of Earth occultation (ideally, these should be consistent). The Earth-occulted background count rate in the range $15-40$\keV\ ($0.194\pm0.002$) was found to be lower than the simulated NXB ($0.271\pm0.004$), hence the Earth-occulted background was adopted for the non X-ray component. This was combined with a cosmic X-ray background (CXB) spectrum that was modelled using a flat response for the PIN. The good detection was made of Mrk~335 in the energy range $15-40$\keV\ using the PIN detector with an effective exposure of 131\ks\ in 2006 and 254\ks\ in 2013. The PIN spectra were fit simultaneously with those from the XIS, with the PIN model enhanced by a factor of 1.18 when the source was in the HXD-nominal position and 1.16 for the XIS-nominal position to account for the cross-calibration of the instruments.

Sufficient photon counts were detected by the PIN to constrain the shape of the broadband X-ray spectrum, in particular the slope of the power law continuum and the reflection fraction, although the PIN data are insufficient to fully constrain spectral features above 10\keV\ such as the Compton hump.

\subsection{NuSTAR}

\textit{NuSTAR} observed Mrk~335 simultaneously with part of the 2013 \textit{Suzaku} observation, offering spectral coverage from 3 to 50\keV. Event lists from \textit{NuSTAR} were reprocessed using the latest calibration following the standard procedure with \textsc{nustardas} v1.4.1. The source spectra were extracted from regions 60\,arcsec in diameter from both the FPMA and FPMB detectors, centred on the point source and corresponding background spectra were extracted from a region of the same size from another region of the same detector. Response matrices were produced with the \textsc{nuproducts} pipeline used to extract the spectra. The spectra from the two observation IDs were summed under average response matrices and the separate (summed) spectra from the FPMA and FPMB detectors were fit simultaneously. Mrk~335 was well detected in the energy range $3-50$\keV.

Preliminary analysis of this spectrum, found that it could not be fit simultaneously with the full \textit{Suzaku} observation, with there being a discrepancy between the $15-40$\keV\ part of the spectrum measured by the \textit{NuSTAR} FPMA and FPMB detectors and the \textit{Suzaku} PIN. The \textit{NuSTAR} spectra suggest a steeper continuum The slope of the X-ray continuum is found to be highly variable throughout the duration of the 2013 observation, hence the combined spectrum over the full period represents the average properties of the source. Since the \textit{NuSTAR} observation was simultaneous with only part of the \textit{Suzaku} observation, it reasonably shows different average properties, hence we do not use the \textit{NuSTAR} data to further constrain parameters derived from the \textit{Suzaku} observation. 

We therefore fit the \textit{NuSTAR} FPMA and FPMB spectra simultaneously with the concurrent part of the \textit{Suzaku} observation (Obs. ID 708016010), with total exposure 59\ks\ and beginning 168\ks\ from the start of the \textit{Suzaku} observation.  A thorough analysis of the complete set of 2013 \textit{NuSTAR} observations, including the 200\ks\ section following the \textit{Suzaku} observation, is conducted by \citet{parker_mrk335}.

\begin{table*}
\begin{minipage}{170mm}
\centering
\caption{\label{obs.tab}X-ray observations of Mrk~335 used in this work to understand the evolution of the X-ray emitting corona. Fluxes are shown to compare the epochs and are calculated from the best-fitting model to the EPIC pn spectrum in the case of \textit{XMM-Newton} and to the combined front-illuminated XIS0 and XIS3 spectra in the case of \textit{Suzaku}. $^\dagger$\textit{NuSTAR} fluxes are measured over the energy range $3-50$\keV\ rather than $0.5-10$\keV\ as used for \textit{XMM-Newton} and \textit{Suzaku}.}
\def\arraystretch{1.5}
\begin{tabular}{llccccl}
  	\hline
   	\textbf{Telescope} & \textbf{Flux State} & \textbf{Obs. ID} & \textbf{Start Date} & \textbf{Exposure} & \textbf{Flux (0.5-10\,keV)} & \textbf{Reference} \\
	\hline
	\textit{XMM-Newton} & High & 0306870101 & 2006-01-03 & 133\ks & $4.08\times 10^{-11}$\ergpcmsqps & \citet{grupe+07} \\
	& Low & 0510010701 & 2007-07-10 & 22.6\ks & $4.53\times 10^{-12}$\ergpcmsqps & \citet{longinotti+08} \\
	& Intermediate & 0600540501 & 2009-06-13 & 82.6\ks & $8.82\times 10^{-12}$\ergpcmsqps & \citet{grupe+12}\\
	& & 0600540601 & 2009-06-11 & 132\ks & $6.91\times 10^{-12}$\ergpcmsqps & \citet{grupe+12}\\
	\hline
	\textit{Suzaku} & High & 701031010 & 2006-06-21 & 151\ks & $3.15\times 10^{-11}$\ergpcmsqps & \citet{larsson+08}\\
	& Low & 708016010 & 2013-06-11 & 119\ks & $4.24\times 10^{-12}$\ergpcmsqps & \citet{gallo+14} \\
	&  & 708016020 & 2013-06-14 & 130\ks & $3.43\times 10^{-12}$\ergpcmsqps & \citet{gallo+14} \\
	\hline
	\textit{NuSTAR}$^\dagger$ & Low & 701031010 & 2013-06-13 & 21\ks & $1.02\times 10^{-11}$\ergpcmsqps & \citet{parker_mrk335}\\
	&  & 708016010 & 2013-06-13 & 22\ks & $1.22\times 10^{-11}$\ergpcmsqps & \citet{parker_mrk335} \\
	\hline
\end{tabular}
\end{minipage}
\end{table*}

\section{Measuring the X-ray emitting corona} 
\subsection{The accretion disc emissivity profile}
\label{emissivity.sec}
Recent detailed analysis of the profile of relativistically broadened emission lines seen in the reflection spectra of bare Seyfert galaxies has revealed the illumination pattern of the accretion disc by the coronal X-ray source, that is the \textit{emissivity profile}. By decomposing the relativistically-blurred reflection spectrum, most notably the prominent iron K$\alpha$ line at 6.4\keV, into the contributions from successive radii in the accretion disc, \citet{1h0707_emis_paper} find that the emissivity profile of the disc in the NLS1 galaxy 1H\,0707$-$495 approximately takes the form of a twice broken power law, falling off steeply with index $>7$ over the inner regions of the disc, then flattening to almost a constant between $5\sim 35$\rg\ before falling off slightly steeper than $r^{-3}$ over the outer part of the disc, the form that is expected theoretically for illumination of an accretion disc in the curved spacetime around a black hole by a coronal X-ray source \citep{miniutti+03,suebsuwong+06}. A similar emissivity profile, flattened between $5\sim10$\rg\ was found by \citet{iras_fix} for the accretion disc in the NLS1 galaxy IRAS\,13224$-$3809.

\begin{figure}
\centering
\subfigure[Isotropic point source] {
\includegraphics[width=80mm]{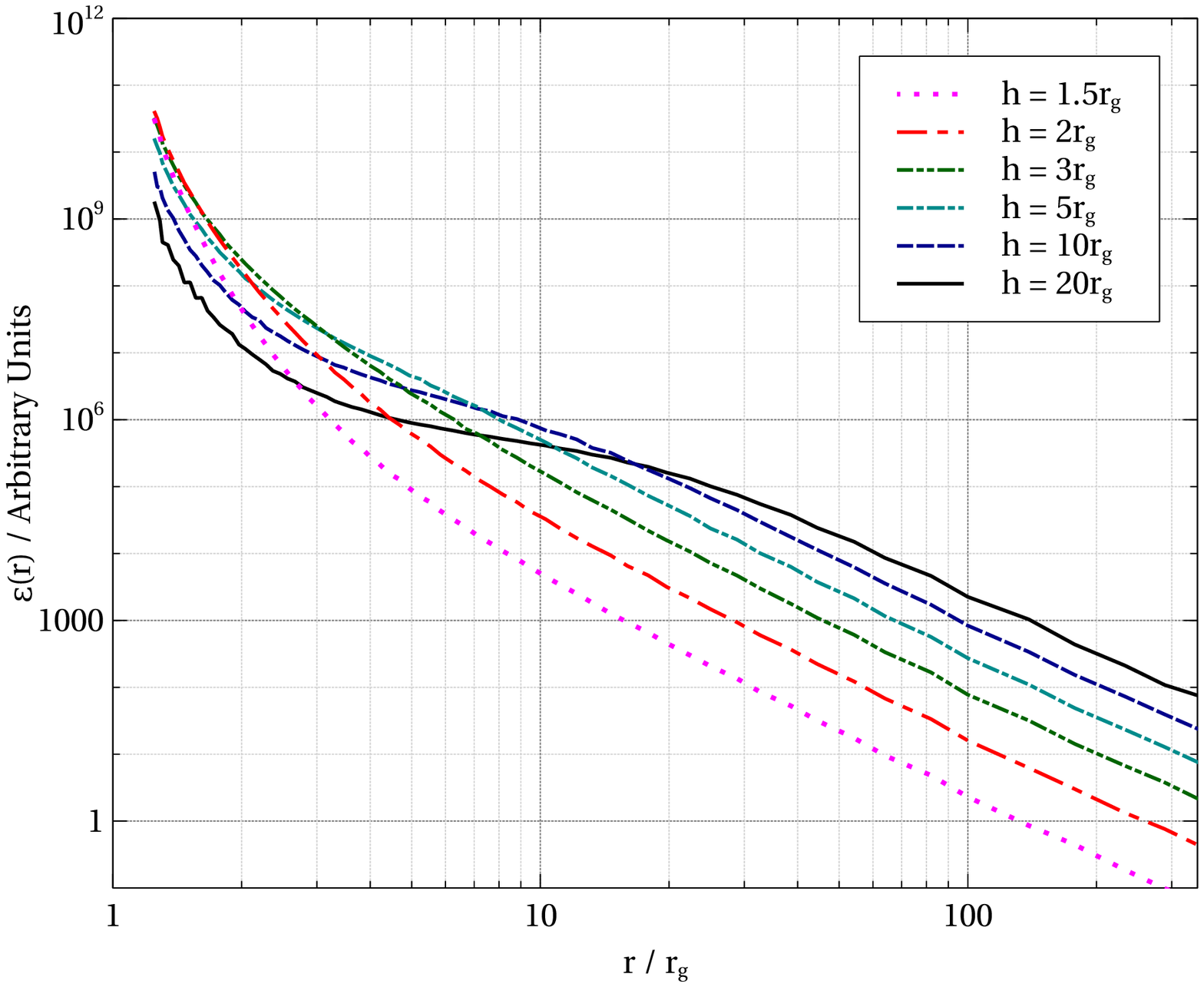}
\label{emis_theory.fig:point}
}
\subfigure[Radially extended coron\ae] {
\includegraphics[width=80mm]{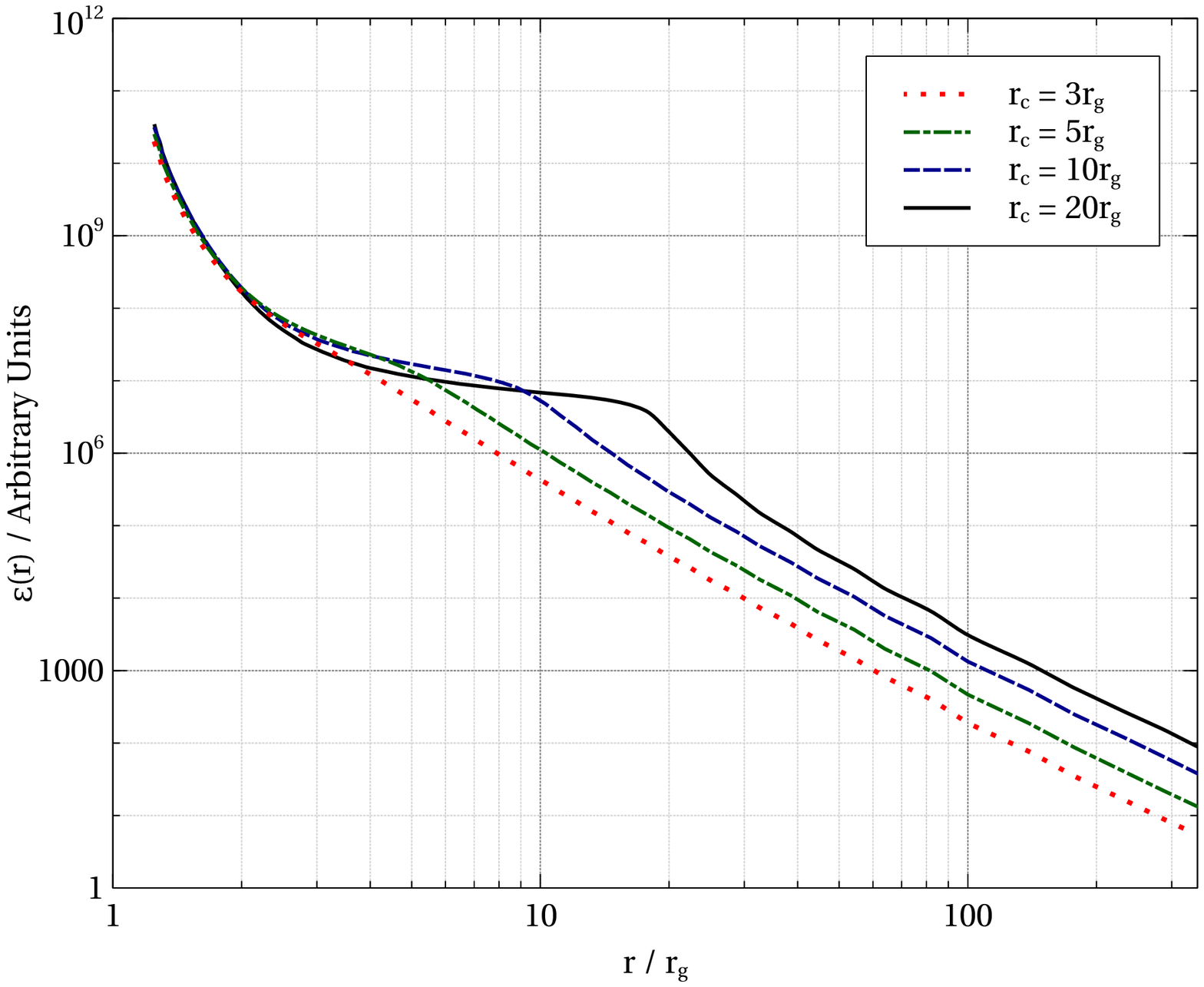}
\label{emis_theory.fig:ext}
}
\subfigure[Vertically extended, jet-like coron\ae] {
\includegraphics[width=80mm]{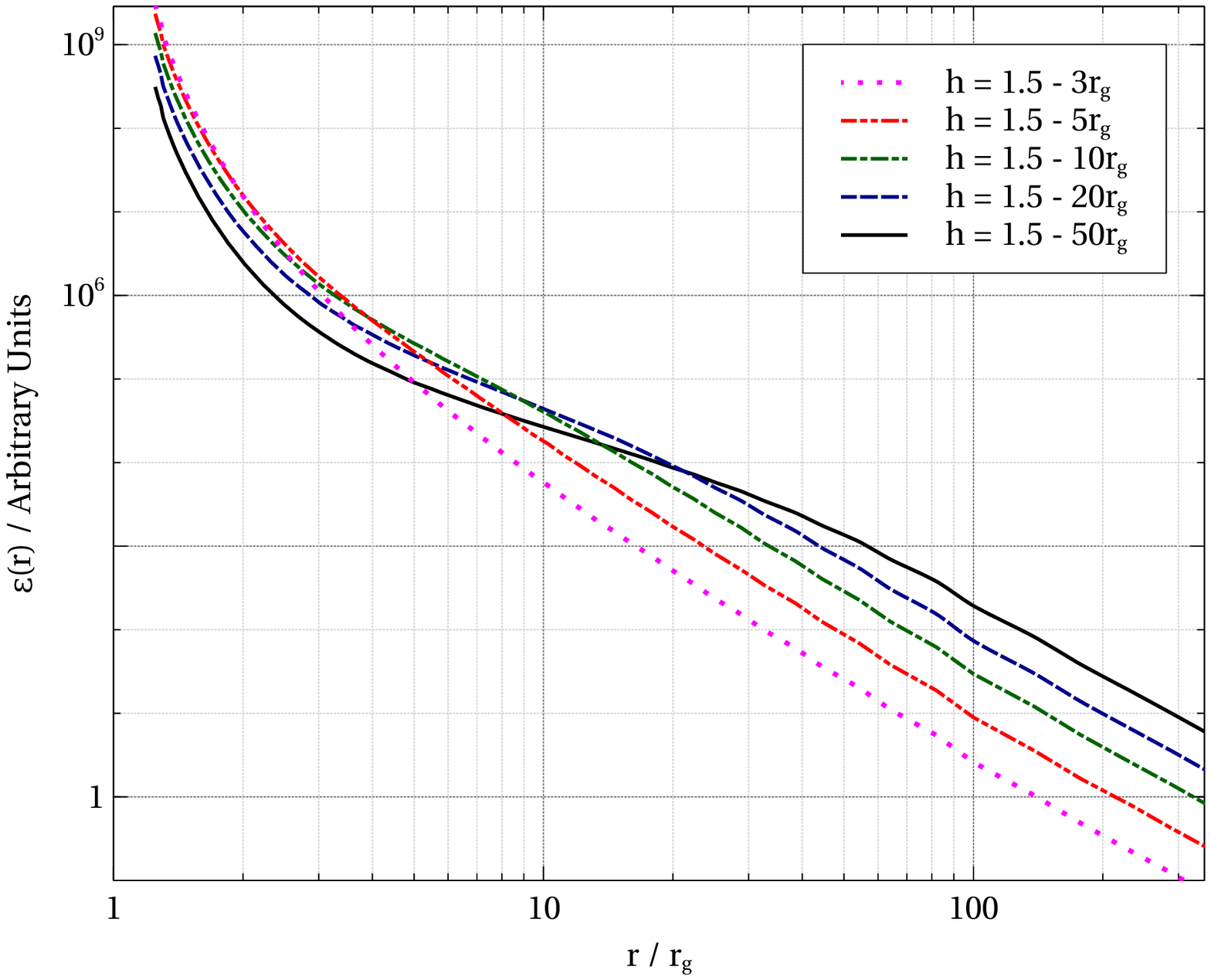}
\label{emis_theory.fig:jet}
}
\caption[]{Theoretical emissivity profiles calculated from general relativistic ray tracing simulations \citep{understanding_emis_paper} for \subref{emis_theory.fig:point} isotropic point sources at varying height above the accretion disc and coron\ae\ extending \subref{emis_theory.fig:ext} radially over the plane of the accretion disc and \subref{emis_theory.fig:jet} vertically, in a jet-like configuration for a maximally rotating Kerr black hole.}
\label{emis_theory.fig}
\end{figure}

\citet{understanding_emis_paper} present a systematic analysis of the expected emissivity profiles for accretion discs illuminated by a range of point-like and extended coron\ae, derived from general relativistic ray tracing simulations. Fig.~\ref{emis_theory.fig} shows predicted emissivity profiles for accretion discs illuminated by a number of different coronal geometries. In the case of an isotropic point source illuminating the disc, the emissivity profile takes the form of a twice-broken power law with the outer break radius approximately coinciding with the height of the source above the disc plane, however with only a slight steeping of the inner disc emissivity profile (with power law indices typically less than 5) until the source is located closer than 3\rg\ to the black hole, at which point, the emissivity profile becomes simply a once-broken power law, falling off progressively more steeply over the inner disc as the source gets closer to the black hole, and tending to $r^{-3}$ at large radius.

In order to explain both the steeply-falling inner disc emissivity profile and the location of the outer break radius in the measured emissivity profile $\gtrsim 10$\rg, \citet{understanding_emis_paper} show that a spatially-extended corona that at least partially covers the inner part of the accretion disc is required to illuminate the disc. The outer break radius of the emissivity profile corresponds to the outermost radial extent of the corona over the plane of the disc, while, in order to reproduce the steep inner part of the profile, the lower bound of the corona must be located less than a few gravitational radii above the plane of the disc. Thus, in the case of a steep inner emissivity profile, the outer break radius reveals the radial extent of the corona over the accretion disc while being relatively insensitive to its vertical extent above the plane of the disc.

Finally, for a corona extended vertically in a jet-like configuration perpendicular to the plane of the accretion disc (\textit{i.e.} whose vertical extent is greater than its radial extent), the emissivity again falls off steeply over the inner part of the accretion disc but then, rather than flattening, is proportional to $r^{-2}$ (as successively higher parts of the jet each contribute an outer break radius according to their height, as in the case of a point source, but contribute less reflected flux getting further from the disc). There is then a slight curvature in the emissivity profile, steepening to $r^{-3}$ at a radius coinciding with the maximum extent of the jet.

\subsubsection{Direct measurement of the emissivity profile}
\label{measure_emis.sec}
The emissivity profile of the accretion disc can be measured directly, making no prior assumption of its form, using the method of \citet{1h0707_emis_paper}, dividing the blurred reflection spectrum into the contributions from successive radii in the disc, described by the \textsc{reflionx} model, convolved with the \textsc{kdblur} blurring kernel in which the inner and outer radius parameters are set accordingly for each annulus and each has a flat emissivity profile. The inclination of the accretion disc (for all annuli) is set to the best fitting value found in the previous fit to the full spectrum, so too are the iron abundance and ionisation parameter of the accretion disc. Also included in the spectral model is any unblurred reflection from distant material as well intrinsic absorption, again with parameters set to the best-fitting values determined prior to measuring the emissivity profile. The normalisation (\textit{i.e.} the contribution) of each annulus to the reflection spectrum is found by minimising $\chi^2$, fitting this model to spectrum over the 3-10\,keV energy range, dominated by the prominent iron K$\alpha$ emission line. In order to constrain the emissivity profile of the inner part of the disc, particularly over the range 5-20\rg, it is necessary to also fit over the range 3-5\,keV, thereby excluding the core of the line dominated by the outer parts of the disc, $> 20$\rg\ (see \citealt{1h0707_emis_paper} for a full discussion).

\subsubsection{Fitting the emissivity profile}
\label{emis_fit.sec}
The best method to determine the emissivity profile of the accretion disc and, hence, the extent of the X-ray emitting corona would be to measure the profile directly, decomposing the reflection spectrum into the contributions from successive radii in the disc, following the method of \citet{1h0707_emis_paper}. Such a decomposition, however, requires a good signal-to-noise detection of the iron K$\alpha$ line, which depends upon the exposure time of the observation as well as the inherent strength of the line; the overall reflected flux from the accretion disc, the iron abundance and ionisation state of the disc all contribute. When we wish to study the evolution of geometry of the corona as the source varies between higher and lower levels of flux during the course of an observation, these required long exposure observations are not available from which the emissivity profile can be measured directly with no prior assumption of its form. In these instances, we are guided by measurements we have from long observations and theoretical predictions and the takes the form of either a once- or twice-broken power law and we fit a model of the relativistically blurred reflection spectrum in which the slopes and break points of the power law forms of the emissivity profile are free parameters.

Firstly, in order to determine whether the once-broken power law for a compact source close to the black hole, or a twice-broken power law for a more extended source yields the better description of the data, it is important to understand how a once-broken power law can be fit to the latter case of a twice-broken power law emissivity profile. To this end, observations of the relativistically blurred reflection of a power law continuum with various twice-broken power law emissivity profiles were simulated using the \texttt{fakeit} command in \textsc{xspec} \citep{xspec}. Data were simulated to a quality comparable to the long observations currently available of NLS1 galaxies; the spectral model and count rate were based upon those found for 1H\,0707-495 \citep{zoghbi+09}. A model consisting of the power law continuum and blurred reflection with a once-broken power law emissivity profile was fit to the simulated spectra. In all cases, the real emissivity profile is guided by the measured profile of 1H\,0707-495 and falls off as $r^{-7}$ over the region $r<5$\rg then flattens to a power law index of zero. The outer break radius is variable and the profile falls of as $r^{-3.3}$ over the outer disc. Results are shown in Table~\ref{plemis.tab}.

\begin{table}
\centering
\caption{The best-fit once-broken power law emissivity profiles to simulated reflection spectra with twice-broken power law emissivity profiles with $q_\mathrm{in} = 7$, $r_\mathrm{br,in} = 5$\rg, $q_\mathrm{mid}=0$ and $q_\mathrm{out} = 3.3$.The outer break radius, $r_\mathrm{br,out}$, of the real emissivity profile is variable.}
\begin{tabular}{cccc}
  	\hline
	Real Profile & \multicolumn{3}{c}{Best Fit Once-Broken Profile} \\
   	$r_\mathrm{br,out}$ & $q_\mathrm{in}$ & $r_\mathrm{br}$ & $q_\mathrm{out}$ \\
	\hline
	7\rg & 7.82 & 4.98\rg & 3.26 \\
	10\rg & 7.81 & 4.35\rg & 3.04 \\
	15\rg & 7.82 & 4.01\rg & 2.69 \\
	20\rg & 7.76 & 3.85\rg & 2.49 \\
	30\rg & 7.66 & 3.67\rg & 2.23 \\
	50\rg & 5.09 & 5.09\rg & 1.73 \\
	\hline
\end{tabular}
\label{plemis.tab}
\end{table}

We find that in all cases, the once-broken power law still provides a good fit to the data ($\chi^2/\nu<1.1$ in all cases).The inner part of the once-broken power law emissivity profile is used to reproduce, almost exactly, the steep inner part of the real emissivity profile, with the inner power law index and break point matching closely to the inner index and first break point of the real emissivity profile. This behaviour can be explained in terms of the majority of the reflected flux (as much as 60 per cent of what is measured by an observer at infinity) arising from these inner regions of the accretion disc \citep{1h0707_emis_paper}. The outer power law index of the once-broken emissivity profile is then used to reproduce the reflected emission from the middle and outer parts of the disc. The measured outer emissivity index is essentially a weighted average of the indices of the real (flattened) middle and outer parts of the profile, with the measured outer emissivity index decreasing as the outer break point moves to larger radius.

In reality, we expect the power law index of the emissivity profile to be $\gtrsim 3$ on the outermost parts of the disc, with steeper profiles being measured for the most compact coron\ae. We therefore conclude that measuring a once-broken power law emissivity profile with a steep inner emissivity index and outer emissivity index less than 3 implies an extended corona a few gravitational radii above the plane of the disc (to give the steep inner index) and extending radially $\gtrsim 10$\rg\ over the inner part of the disc. Given such a result, a model reflection spectrum with a twice-broken power law emissivity profile may then be fit to the data to determine the location of the outer break radius, though since such a model has more free parameters than the once-broken power law model, constraining these parameters may requirer higher quality data.

\subsection{The reflection fraction}
In a scenario in which the geometry or spatial extent of the corona is varying, it should be expected that the relative fraction of photons that are detected in the power law continuum and in the reflection changes. As the X-ray source becomes confined to a more compact region around the black hole, gravitational light bending causes more of the photons emitted from the corona to be focussed towards the black hole and hence on to the inner regions of the accretion disc, rather than being able to escape in order to be detected as part of the X-ray continuum. As such, the fraction of reflected photons relative to continuum photons is enhanced.

The reflection fraction is not, na\"ievly expected to drop below unity. If an isotropic X-ray source is located above an infinite accretion disc subtending solid angle $2\pi$ to an observer at the source location, half of the emitted continuum photons will hit the disc while half will be able to escape to form the continuum.

Such a picture was invoked by \citet{miniutti+04} to explain the relative constancy of the reflected flux in the NLS1 galaxy MCG--6-30-15 while the continuum flux is seen to vary greatly. They calculate the variation in reflected and continuum flux for a constant luminosity, isotropic point source of X-rays that moves up and down the rotation axis of the black hole and show that this scenario is consistent with the observed variation in fluxes. Likewise, \citet{1h0707_jan11} show that the low flux state that 1H\,0707$-$495 was seen to drop in to in January 2011, in which solely the relativistically blurred reflection spectrum from the accretion disc was seen with little or no contribution to the spectrum from the directly-observed power law continuum, can be understood in terms of the previously extended corona collapsing down to a confined region spanning just a few gravitational radii around the central black hole, meaning almost all of the continuum photons are focussed onto the disc or lost beyond the event horizon.

Ray tracing simulations that count the number of rays emitted from a corona that are able to escape to infinity (to be detected as the X-ray continuum) and that are incident upon the disc, following \citet{understanding_emis_paper} and \citet{1h0707_jan11}, allow the reflection fraction to be predicted. When photons are reflected from an accretion disc upon which the majority of photons fall on the innermost few gravitational radii (as suggested by the measured emissivity profiles), ray racing simulations show that less than half of the reflected photons are able to escape \citep{understanding_emis_paper}. Many will either return to the disc under the strong gravitational field around the black hole to give rise to second (and higher) order reflections or will be lost beyond the black hole event horizon which will reduce the measured reflection fraction. Thus, in order to accurately predict the reflection fraction from a given corona, it is necessary to compute the fraction of the reflected photons that will be observed which will, in turn, depend upon the measured emissivity profile. However, given this information, the reflection fraction can be used to place additional constrains on the corona, including its vertical extent that is not well constrained by the emissivity profile alone. Alternatively, measuring the variation in the reflection fraction over time for a given source shows how the extent of the corona is \textit{changing}.

\section{Long timescale variability}
In order to understand the long timescale variability of Mrk~335 in terms of changes to the X-ray emitting corona, the spectrum from each of the six observations with \textit{XMM-Newton} and \textit{Suzaku} were analysed independently using \textsc{xspec} \citep{xspec}. 

\subsection{The spectral model}
The spectra were fitted with models in \textsc{xspec} based upon power law continuum emission observed directly from the corona and the relativistically blurred reflection thereof from the accretion disc. The rest frame reflection spectrum from the accretion disc is modelled by the \textsc{reflionx} code of \citet{ross_fabian} and is convolved with the profile of a relativistically blurred emission line with either a once- or twice-broken power law accretion disc emissivity profile using the \textsc{kdblur2} model.

Reflection of the X-ray continuum by distant material, producing, among other features, a narrow iron K$\alpha$ emission line on top of the broad line at 6.4\keV\ is accounted for by including a second, unblurred \textsc{reflionx} component with low ionisation parameter. Principal component analysis (PCA) conducted by \citet{gallo+14} between the 2006 high flux and 2013 low flux observations with \textit{Suzaku} shows a narrow component at 6.4\keV\ in the first principal component, indicating variation in the distant reflector on timescales of years. To account for this, the flux of the distant reflection component is allowed to vary as a free parameter in the fits to the spectra obtained during each epoch.

Mrk~335 shows variable intrinsic absorption in its X-ray spectrum with \citet{longinotti+13} reporting the discovery of an ionised wind outflowing at around 5000\kmps. High resolution X-ray spectra obtained with the \textit{XMM-Newton} Reflection Grating Spectrometer (RGS) revealed that this wind is composed of three distinct components in different ionisation states ($\xi\sim10$, 100 and 1000\ergcmps where the ionisation parameter for the illumination of material with atomic hydrogen number density $n$ illuminated by ionising flux $F$, $\xi = 4\pi F / n$). All three absorbers are clearly detected in the 2009 intermediate state \textit{XMM-Newton} observations \citep{longinotti+13,gallo+13}, while to explain the \textit{XMM-Newton} high flux state spectrum of 2006, only the lesser ionised absorber is required, likewise in the \textit{XMM-Newton} low flux observation. On the other hand, the low flux state of Mrk~335 recorded by \textit{Suzaku} in 2013 shows evidence of only the hottest, most ionised of these absorbers \citep{gallo+14} while the high flux spectrum observed by \textit{Suzaku} in 2006 requires no intrinsic absorption \citep{larsson+08}. The three ionised, warm absorbers are modelled by pre-calculated tables computed by \citet{longinotti+13} using \textsc{xstar} and the appropriate combination of tables is included in the model for each of the observations along with the absorption by martial in our own galaxy along the line of sight, described by the \textsc{tbabs} model. The column density of hydrogen atoms through our Galaxy to Mrk~335 is $3.6\times 10^{20}$\psqcm.

In addition to the intrinsic absorption, the low flux epochs also show evidence for ionised emitting material and we find that additional line emission at 0.88\keV. This is modelled by a narrow Gaussian profile ($\sigma=1$\eV) with centroid energy and normalisation fit to the observed spectrum as free parameters.

\begin{figure*}
\centering
\subfigure[\textit{XMM-Newton} High flux (2006 January)] {
\includegraphics[width=80mm]{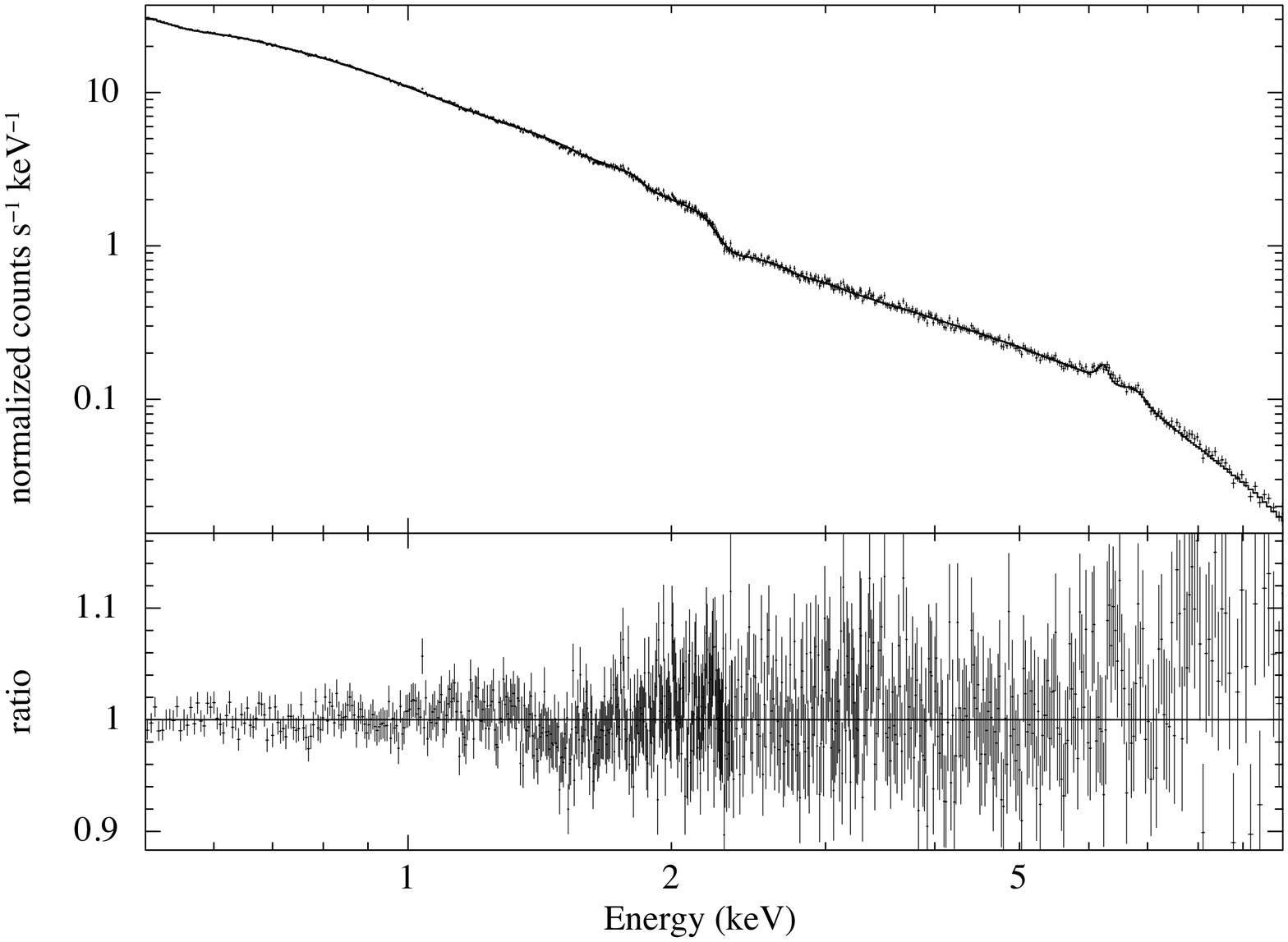}
\label{fit_spectra.fig:xmm_high}
}
\subfigure[\textit{XMM-Newton} Intermediate flux (2009 June)] {
\includegraphics[width=80mm]{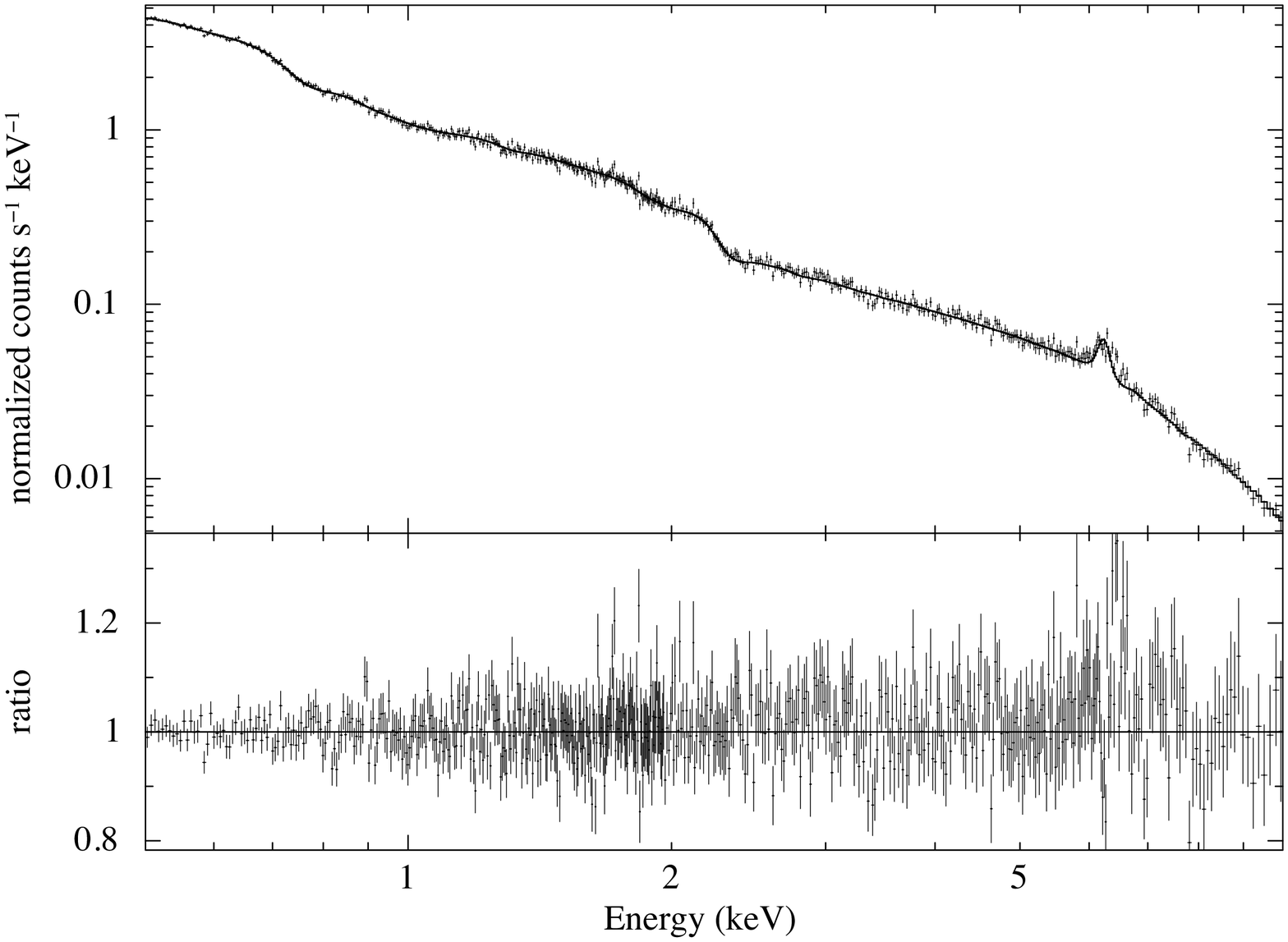}
\label{fit_spectra.fig:xmm_intermediate}
}
\subfigure[\textit{XMM-Newton} Low flux (2007 October)] {
\includegraphics[width=80mm]{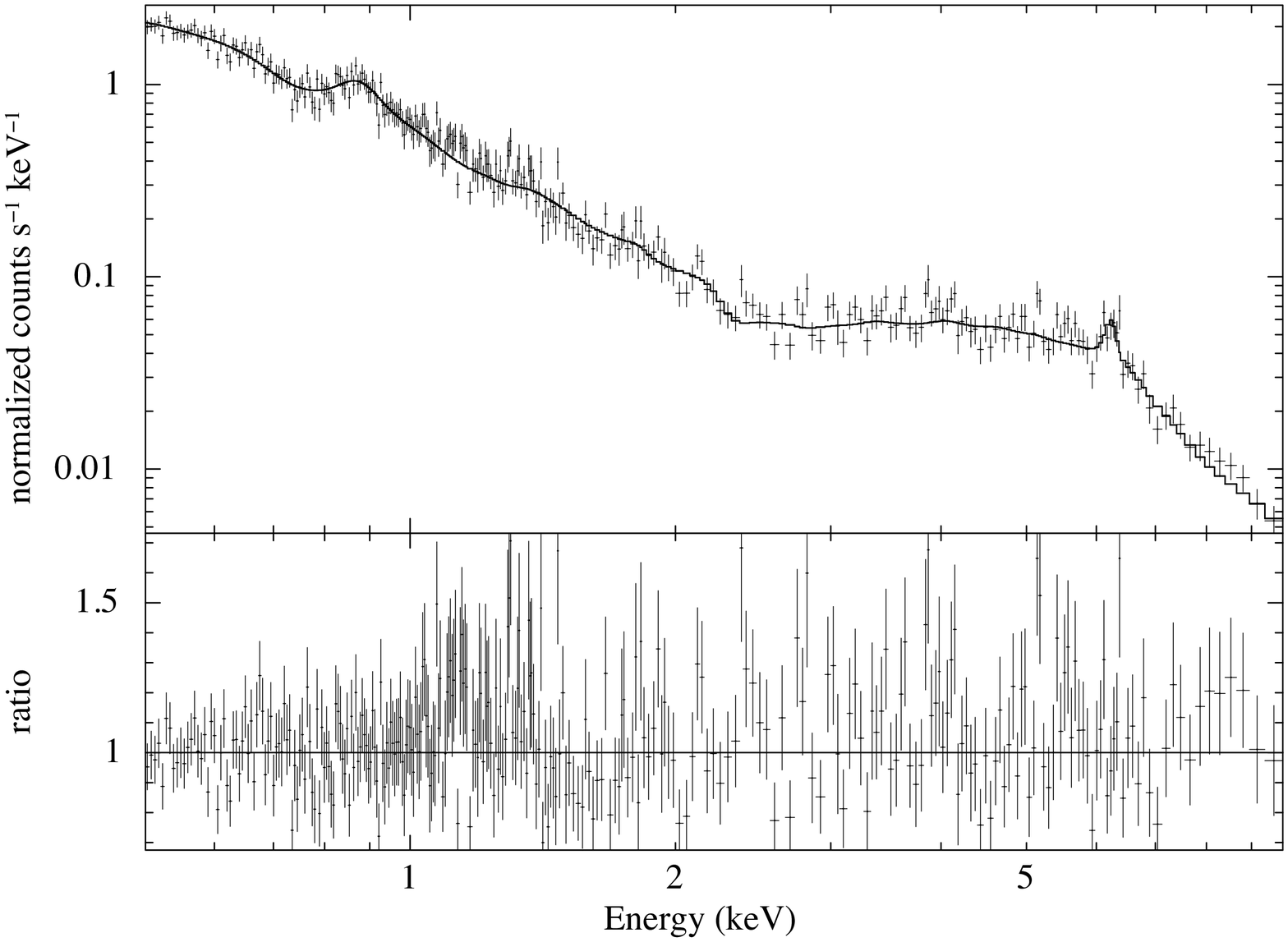}
\label{fit_spectra.fig:xmm_low}
}
\subfigure[\textit{Suzaku} High flux (2006 June)] {
\includegraphics[width=80mm]{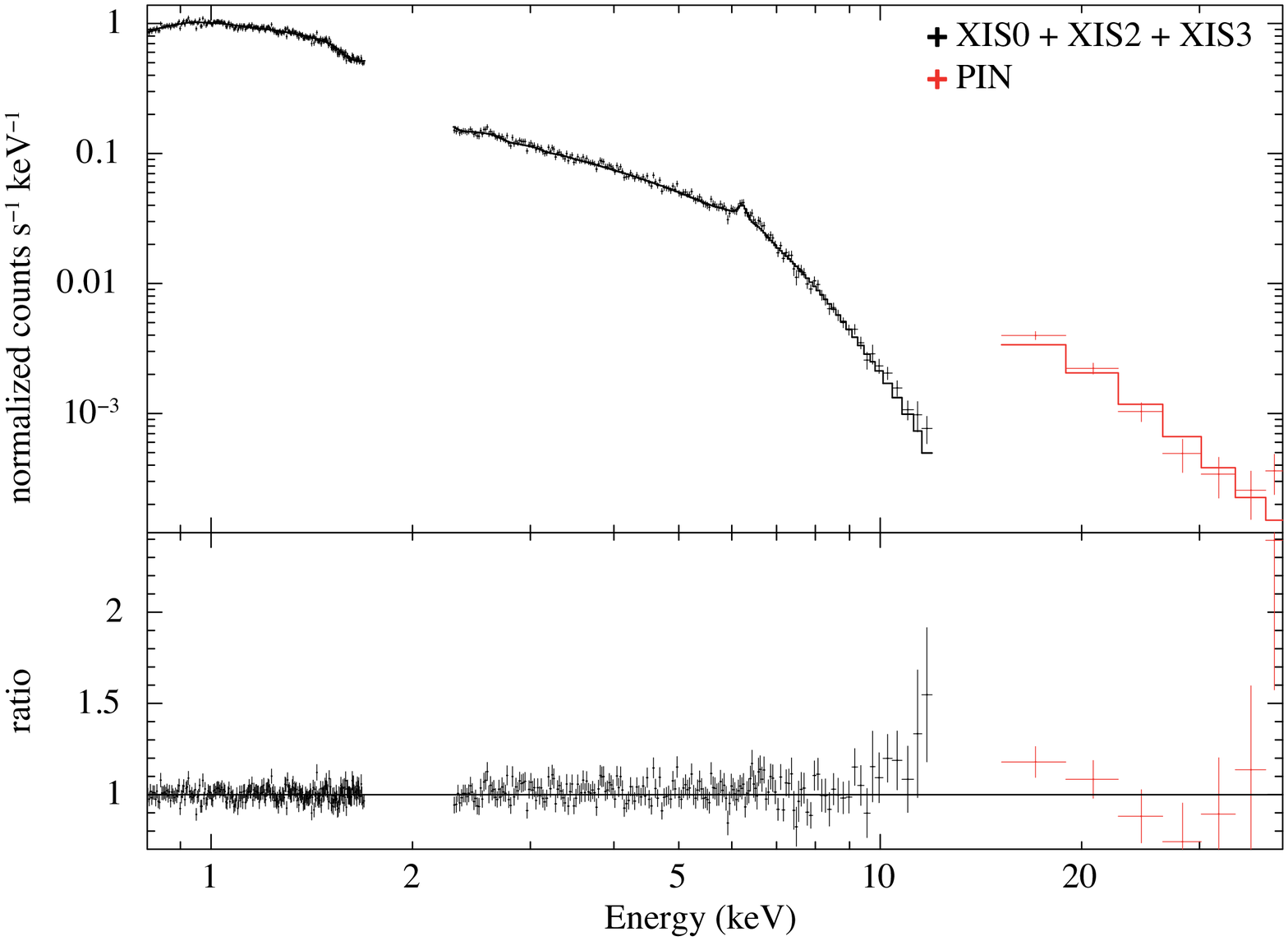}
\label{fit_spectra.fig:suzaku_high}
}
\subfigure[\textit{Suzaku} Low flux (2013 August)] {
\includegraphics[width=80mm]{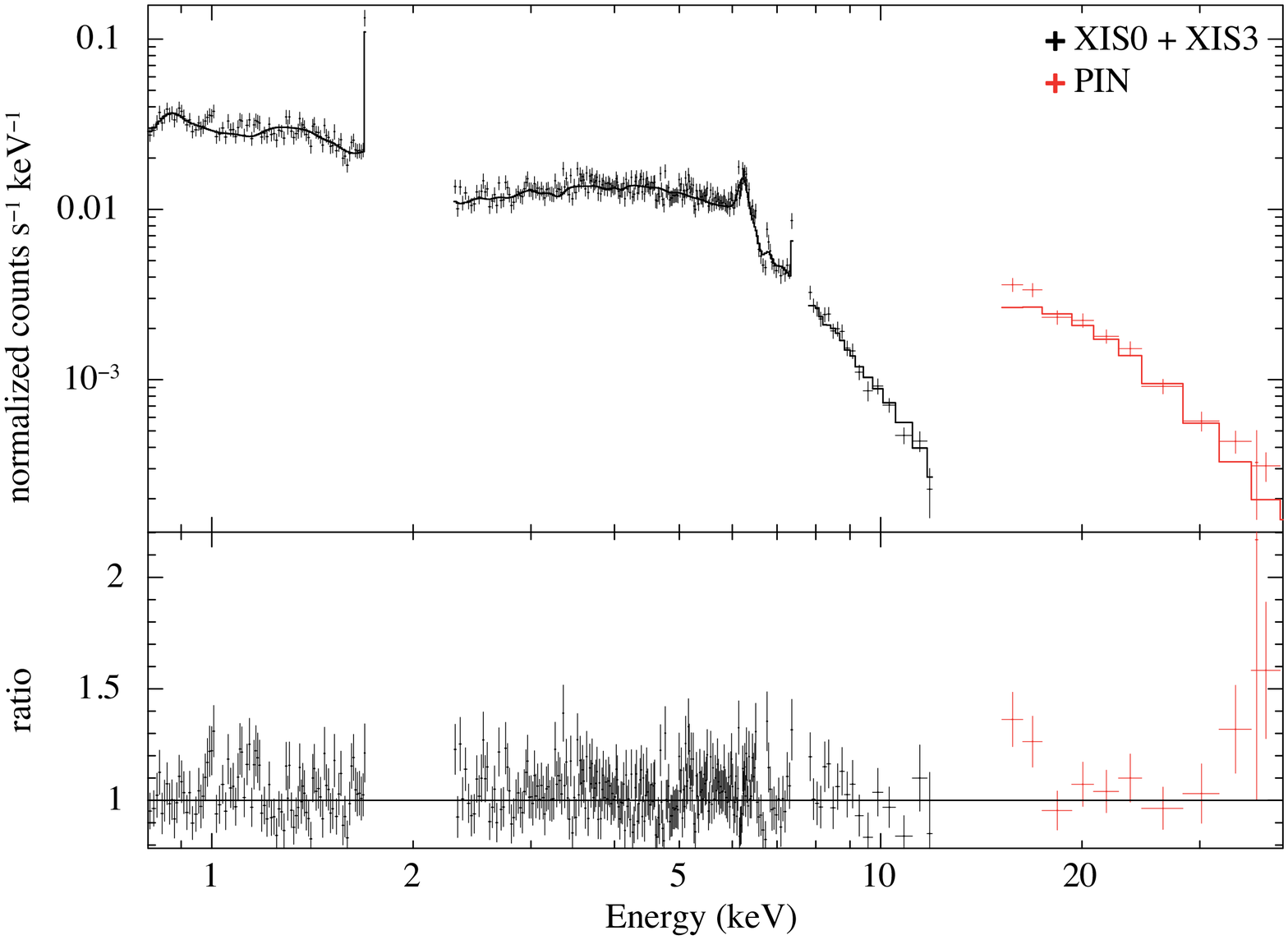}
\label{fit_spectra.fig:suzaku_low}
}
\subfigure[Base Model] {
\includegraphics[width=82mm,trim=-2mm 0 0 0]{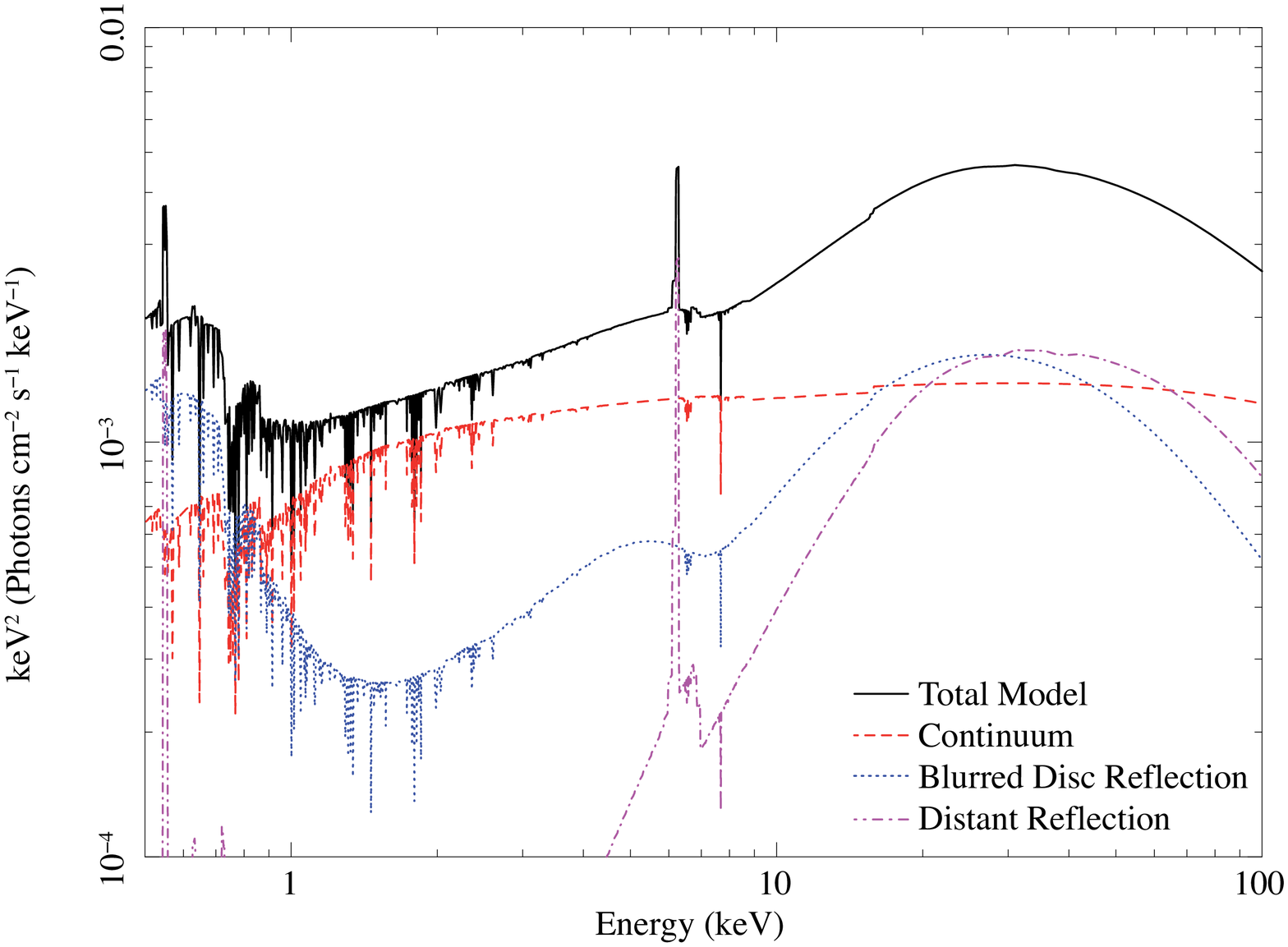}
\label{fit_spectra.fig:model}
}
\caption[]{The spectra of Mrk~335 during the observed epochs of varying flux with the best-fitting models consisting of power law continuum emission from the corona, the relativistically blurred reflection of this from the accretion disc, unblurred reflection from distant material and intrinsic absorption by outflowing material, appropriate for each epoch. The parameters derived from these preliminary fits are used in the base model to measure the emissivity profile of the accretion disc. The residuals apparent around the blue-shifted edge of the 6.4\keV\ iron K$\alpha$ line in \subref{fit_spectra.fig:xmm_intermediate} and \subref{fit_spectra.fig:suzaku_high} arise due to inadequacies in the modelling of the accretion disc emissivity profile and are removed once the emissivity profile is fit explicitly (Fig.~\ref{emissivity_epochs.fig}), as illustrated in \citet{1h0707_emis_paper}. Also shown in \subref{fit_spectra.fig:model} is the spectral model used to determine the basic parameters of the X-ray continuum and reflection spectrum. The model is shown with the best fitting parameters to the 2009 intermediate flux epoch during which the most significant absorption was detected.}
\label{fit_spectra.fig}
\end{figure*}

The spectra obtained during each of the observed epochs with the respective best-fitting models are shown in Fig.~\ref{fit_spectra.fig} along with the basic model fit to the data.

\subsection{Measuring changes in the corona}
The parameters of interest in understanding the changes in the geometry and energetics of the corona are the reflected and continuum fluxes (and the ratio thereof), the photon index of the power law continuum and the emissivity profile of the accretion disc. The above model was initially fit to the observed spectra during each of the epochs, incorporating the appropriate intrinsic absorption for the flux state in question, in order to find the best-fitting values of the photon index of the continuum, and the inner radius, inclination, iron abundance and ionisation parameter of the accretion disc, as well as the column density and exact ionisation parameters of the intrinsic absorbers. The results of these fits to the  spectra of Mrk~335 during each of the observed epochs are shown in Table~\ref{fit_results_epochs.tab}.

The reflection fraction, $R$, is measured as the ratio of the photon fluxes from the blurred reflection and power law continuum model components, extrapolated over the energy range 0.1-100\keV. These fluxes are measured by applying the \textsc{cpflux} pseudo-model in \textsc{xspec} to the appropriate model components, allowing their fluxes and the errors thereof to be determined as model parameters. For comparison with theoretical predictions of the fraction of rays emitted from the corona that are incident upon the accretion disc and able to escape to be observed as part of the continuum, it is important to consider as wide an energy range as possible in the reflection fraction. Ray tracing calculations are independent of the model reflection spectrum and do not account for the energy at which photons incident on the disc will be re-emitted in the reflection spectrum, thus a comparison requires an estimate of the total number of reflected photons. It is also possible to compute reflection fractions over a more restricted energy range, such as 20-40\keV, encompassing the Compton hump. While insensitive to variations in the emission lines below 1\keV\ that are dependent on the ionisation state, we find that the reflection fraction in this energy band is highly sensitive to the photon index of the primary continuum and is, hence, not a reliable estimate of the total fraction of the X-ray continuum that is reflected. We, however, include the reflection fraction over the 20-40\keV\ energy band for comparison with more recent modelling of accretion disc reflection spectra, for instance that conducted by \citet{parker_mrk335}. It should be noted that 0.1-100\keV\ energy band is not without degeneracies, since the reflected flux in emission lines below 1\keV\ increases as the disc becomes more ionised. When there is spectral coverage to low energies, this can be well constrained. In the case of \textit{Suzaku} XIS spectra, however, the lack of data below 0.8\keV\ leads to some uncertainty in the flux arising from the extrapolated section of the reflection spectrum. It is therefore necessary to allow the ionisation parameter to vary freely when computing the error in the reflection fraction. The quoted uncertainties reflect the range of ionisation parameters admitted by the available data.

The errors in each of the model parameters were determined using the conventional approach in \textsc{xspec}; the standard deviation, $\sigma$, of any model parameter given the data is found by stepping through values of that parameter until the fit statistic, $\chi^2$ increases by 1 from its minimum, after finding the best-fitting value of the other free model parameters for the current value of the parameter of interest. Errors are quoted at the 90 per cent confidence level ($\Delta \chi^2 = 2.706$).

The values determined for these free model parameters were then used to construct a spectral model in order to measure the emissivity profile of the accretion disc, following the method of \citet{1h0707_emis_paper}, detailed in Section~\ref{measure_emis.sec}. The emissivity profiles of the accretion disc during each of the epochs is shown in Fig.~\ref{emissivity_epochs.fig}.

\label{mcmc.sec}
Errors in the emissivity profiles are calculated from Monte Carlo Markov Chain computation of the probability distribution of the normalisation of the reflection contributed by each annulus, given the observed spectrum \citep[see also][on the application of MCMC methods to X-ray spectral fitting]{steiner+12,reynolds+12}. A so-called `walker' is started at a given point in the parameter space, here taken to be the best-fitting normalisation found during the calculation of the emissivity profile. From this point, a random step is taken in the parameter space with distributions in each parameter drawn from the diagonal of the covariance matrix found during the fit (this is the variance of each parameter). The likelihood of the data given the model with these new parameters is then computed. If this is greater than the likelihood at the previous position (i.e. the fit to the data is improved), the walker definitely moves to the new location. If not, the walker moves to the new location, but only with a probability defined by the ratio of the likelihoods at the new and old locations. Therefore, the walker may move or the step may be rejected, causing the walker to stay at the same location, on any iteration. The process is then repeated, with the walker either moving or staying put over a defined number of steps, and the sample of locations visited by the walkers in the parameter space follows the probability distributions of the free parameters.

We employ the MCMC sampler of \citet{goodman_weare} to generate each random step\footnote{We use the \textsc{xspec\_emcee} implementation of the \textsc{python emcee} package for X-ray spectral fitting in \textsc{xspec} by Jeremy Sanders (http://github.com/jeremysanders/xspec\_emcee)}, which, through an affine transformation of the parameter space, is better able to cope with the degeneracy between the normalisations of neighbouring annuli in their contributions to the reflection spectrum (essentially, performing a co-ordinate transformation such that the random steps are taken along the axes of any degeneracy in the space). We trace multiple walkers through the parameter space simultaneously (for good sampling, the number of walkers should be more than twice the number of model parameters) and the final chain is formed by combining the steps of all the walkers. The first 1000 steps of each walker's chain are discarded (or `burned') to remove bias introduced by the choice of starting location, then the probability distributions from which the errors are found are computed by constructing histograms of each of the parameter values across all steps in the chain.

Using MCMC calculations to compute the errors on the emissivity profiles is advantageous in terms of computational efficiency, allowing the confidence limits on the normalisations of the reflection from all 35 annuli to be computed simultaneously, while MCMC methods are less vulnerable to local minima in the goodness-of-fit, with their ergodicity meaning that they explore the full parameter space and also less vulnerable to steep gradients in the goodness-of-fit causing the error calculation to simply peg at the hard limits set for the parameter. 

\subsection{Results}

\subsubsection{Properties of the X-ray spectrum}

Initially, looking at the best-fitting parameters in Table~\ref{fit_results_epochs.tab} of a spectral model consisting of power law continuum emission from the corona around the black hole and the relativistically blurred reflection of this emission from the accretion disc, we find that that as the flux decreases, the reflection fraction increases significantly from a little over unity during the \textit{XMM-Newton} observation in the 2006 high flux epoch to $R\sim 5$ in the 2013 low flux epoch observed by \textit{Suzaku}. During the 2006 high flux epoch observed with \textit{Suzaku}, the reflection fraction decreased dramatically to $R\sim 0.26$, notably less than unity as would be expected classically in the case of a corona illuminating an accretion disc extending to infinity in the absence of gravitational light bending and other relativistic effects. We see that the flux of the unblurred reflection varies slightly between the epochs as suggested by the principal component analysis of \citet{gallo+14}, though it does not simply follow the high and low flux epochs. If the unblurred reflection arises from material far from the central engine, for instance from the obscuring torus hypothesised in Seyfert galaxies, the reflected flux will vary as the average source luminosity over a timescale corresponding to the range of light travel times to different parts of the reflector.

\begin{table*}
\begin{minipage}{175mm}
\centering
\caption{\label{fit_results_epochs.tab}The best-fitting values of the model parameters used to measure the emissivity profile of the accretion disc during each epoch; the photon index of the X-ray continuum, $\Gamma$, the inclination, $i$, of the normal to the accretion disc to the line of sight and the inner radius of the accretion disc, $r_\mathrm{in}$ as well as the iron abundance, $A_\mathrm{Fe}$ and ionisation parameter, $\xi$, of the material in the accretion disc, and the reflection fraction, defined as the ratio of photon counts between the reflected and power law continuum model components, $R=N_\mathrm{ref}/N_\mathrm{pl}$, firstly extrapolated over the energy range 0.1-100\keV\ and also over the range 20-40\keV\ for comparison with other reflection models. The flux of the unblurred reflection from material distant from the black hole is measured over the 0.1-100\keV\ energy band.}
\def\arraystretch{1.5}
\begin{tabular}{llcccccc}
  	\hline
   	\textbf{Component} & \textbf{Parameter} & \textbf{All} & \textbf{XMM High} & \textbf{Suzaku High} & \textbf{XMM Int.} & \textbf{XMM Low} & \textbf{Suzaku Low} \\
	\hline
	\textsc{powerlaw} & $\Gamma$ & & $2.523_{-0.010}^{+0.011}$ & $2.16_{-0.01}^{+0.02}$ &  $1.90_{-0.02}^{+0.02}$ & $2.36_{-0.08}^{+0.09}$ & $1.91_{-0.07}^{+0.04} $ \\
	\hline
	\textsc{kdblur2} & $i$ / deg & $57.1_{-1.2}^{+0.8}$  & $58.8_{-3.0}^{+1.9}$  & $59_{-3}^{+7}$ & $52_{-14}^{+7}$ & $66_{-1}^{+2}$ & $58_{-6}^{+4}$\\
	& $r_\mathrm{in}$ / \rg & & $1.235^{+0.003}$  & $2.5_{-1.5}^{+2.7}$ & $1.24_{-0.05}^{+0.14}$ & $1.28_{-0.05}^{+0.02}$ & $1.25_{-0.02}^{+0.03}$ \\
	\hline
	\textsc{reflionx} & $A_\mathrm{Fe}$ / solar & $2.6_{-0.2}^{+0.3}$  & $1.75_{-0.17}^{+0.13}$  & $1.9_{-0.1}^{+4.2}$ & $3.6_{-0.7}^{+1.2}$ & $4.2_{-0.6}^{+0.9}$ & $6.7_{-1.4}^{+0.8}$\\
	& $\xi$ / \ergcmps & & $58_{-4}^{+12}$  & $105_{-50}^{+14}$ & $250_{-20}^{+30}$ & $21.0_{-0.8}^{+0.9}$ & $13_{-5}^{+7}$\\
	& $R_{\,0.1-100\,\mathrm{keV}}$ & & $1.3_{-0.2}^{+0.5}$ & $0.26_{-0.02}^{+0.04}$ & $1.8_{-0.3}^{+0.4}$ & $>40$ & $6_{-3}^{+4}$ \\
	& $R_{\,20-40\,\mathrm{keV}}$ & & $3.2_{-0.5}^{+1.0}$ & $0.59_{-0.05}^{+0.05}$ & $1.1_{-0.2}^{+0.2}$ & $>26$ & $35_{-12}^{+12}$ \\
	\hline
	Unblurred refl. & \multicolumn{2}{l}{$F$ / $10^{-3}\,\mathrm{ph}\,\mathrm{cm}^{-2}\,\mathrm{s}^{-1}$} & $1.6_{-0.2}^{+0.2}$ & $1.3_{-0.3}^{+0.3}$ & $0.89_{-0.05}^{+0.10}$ & $0.8_{-0.4}^{+0.4}$ & $1.2_{-0.2}^{+0.2}$ \\
	\hline
	Goodness of fit & $\chi^2 / \nu$ & 1.10 & 1.06 & 1.04 & 1.00 & 1.12 & 1.04 \\
	\hline
\end{tabular}
\end{minipage}
\end{table*}

The photon index was found to vary by as much as 30 per cent between the observed epochs, with the continuum spectrum softening from the low to the high flux epochs, with the photon index increasing from around 1.9 during the \textit{Suzaku} low and \textit{XMM-Newton} intermediate flux epochs to 2.5 during the \textit{XMM-Newton} high flux epoch.  This is in accordance with the general trend seen in AGN \citep{mark_edel_vaughan}. We note, however, that the low flux epoch observed by \textit{XMM-Newton} does not fit this trend, with a photon index of 2.4 measured during this observation. We caution that the directly detected continuum flux is so low during this observation that the continuum photon index is constrained only by the slope of the emission reflected from the accretion disc.

During the 2009 intermediate flux epoch, the X-ray count rate was seen to increase from 2\ctsps\ to 5\ctsps\ during the first orbit of the observation, remaining at this higher level (averaging 4\ctsps) during the second orbit. The changes in the X-ray spectrum as the count rate increased are considered in detail by \citet{gallo+13}. As the X-ray count rate increased, the continuum spectrum was found to soften with the photon index increasing from $1.85_{-0.05}^{+0.05}$ to $1.99_{-0.03}^{+0.04}$, consistent with the behaviour seen between epochs, however variation in the reflection fraction between 0.1 and 100\keV\ as well as the ionisation parameter were found to be within the statistical errors. We find that the total photon count detected during each of the orbits was insufficient to detect changes in the emissivity profile of the accretion disc, hence it is necessary to consider the summed spectrum to make a measurement of the average emissivity profile over the course of these observations.

During all epochs, we find that the maximum measured redshift in the wing of the relativistically broadened iron K$\alpha$ emission line is statistically consistent with the accretion disc extending as far in as the innermost stable circular orbit of a maximally rotating black hole at $r=1.235$\rg\ supporting findings that the black hole spin, $a>0.9$ \citep{gallo+14}. There is no evidence for truncation of the accretion disc between the high and low flux epochs.

It should be noted that although the spectra taken during the different epochs were fitted separately, the inclination of the accretion disc with respect to the line of sight (which should not change between the observations) is found to be consistent within statistical errors between the spectra. While it is possible that the inner radius of the accretion disc varies between the different epochs, identifying the inner edge of the accretion disc with the innermost stable circular orbit (ISCO), the spin of the black hole (which is not expected to change on the timescales between the observations) is consistent with maximal in each case. 

We note, however, that there is a systematic offset between the iron abundances measured from the reflection spectra during each epoch. The elemental abundances within the accretion flow are not expected to vary on the timescale of $\sim 10$ years between these observations. The iron abundances measured from the two \textit{Suzaku} observations are consistent within their respective errors, however there is notable deviation between the \textit{XMM-Newton} high and \textit{Suzaku} low flux epochs. We therefore also fit the model to all of the spectra simultaneously, tying the values of the iron abundance and accretion disc inclination between the spectra. We find the iron abundance to be $2.6_{-0.2}^{+0.3}$ times the Solar value (which is still inconsistent with the \textit{Suzaku} low flux epoch within statistical errors alone).

\citet{ross+02} report that reflected X-rays returning to the accretion disc due to gravitational light bending to be reflected multiple times causes atomic features in the reflection spectrum to be enhanced, mimicking an enhanced iron abundance. During the lower flux epochs, the emissivity profile of the accretion disc is more centrally concentrated. More X-rays being reflected closer to the black hole will increase the returning radiation to the disc and could, hence, explain the apparent enhancement in the iron abundance during the lower flux epochs.

We caution, however, that, particularly in the \textit{Suzaku} spectra, the iron abundance is not particularly tightly constrained and since the \textit{Suzaku} spectra are only fit upwards of 0.8\keV, cutting out much of the iron L line as well as the soft excess, composed of a number of emission lines in the reflection model, the iron abundance is, to some extent, degenerate with the normalisations of the continua and reflection spectra. When the soft excess is not measured, increasing the iron abundance has the same effect as increasing the normalisation of the reflection spectrum with respect to that of the continuum, merely increasing the flux in the iron K$\alpha$ line above the continuum. Uncertainties in the iron abundance do not, however, affect the accuracy of the measured accretion disc emissivity profiles. It can readily be shown that an increased iron abundance improves the statistical constraint on the emissivity measured from the iron K$\alpha$ line, with a greater number of photon counts to constrain the contribution from each part of the disc, but an under- or over-estimate of the iron abundance affects only the overall (arbitrary) normalisation of the emissivity profile, not the measured shape of the profile, in which we are interested.

\begin{figure}
\centering
\includegraphics[width=85mm]{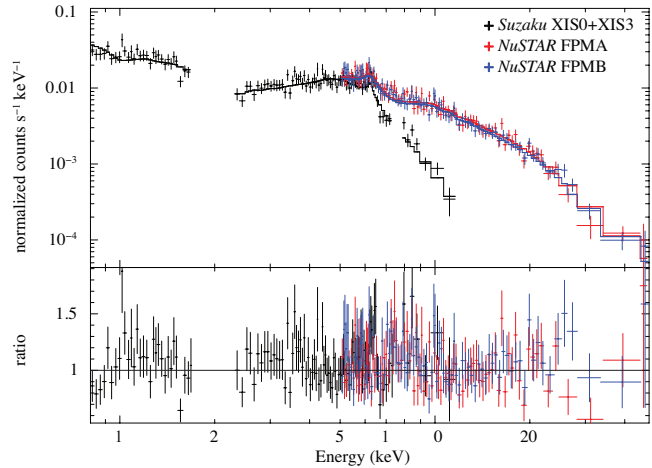}
\caption[]{Simultaneous fit to the concurrent sections of the 2013 \textit{Suzaku} (summed XIS0 and 3) and \textit{NuSTAR} (separate FPMA and FPMB) spectra with the base model ($\chi^2/\nu = 1.05$).}
\label{xis_nustar_simfit.fig}
\end{figure}

Fitting the \textit{NuSTAR} spectra simultaneously with the concurrent part of the \textit{Suzaku} low flux observation to extend energy coverage up to 50\keV\ (Fig.~\ref{xis_nustar_simfit.fig}) resolves this discrepancy. The iron abundance is constrained to $2.4_{-0.5}^{+0.6}$ times the Solar value, in line with the simultaneous fit to all epochs. During this simultaneous section of the 2013 observations, including the \textit{NuSTAR} data also constrains the ionisation parameter of the accretion disc to $\xi < 1.2$, while the X-ray continuum was found during this section to have softened to $\Gamma = 2.50_{-0.08}^{+0.09}$ (the \textit{NuSTAR} observation includes the beginning of the flare seen with \textit{Suzaku} during which the continuum softened). The reflection fraction over the energy band 0.1-100\keV\ as found to be $1.0_{-0.3}^{+0.4}$ (consistent with the low value measured from the \textit{Suzaku} spectrum as the flare begins) though there is no reason to believe that the photon index, reflection fraction or even the ionisation parameter remain constant through the entire 2013 observation given the variability that is observed. The fit to this simultaneous section of the \textit{Suzaku} and \textit{NuSTAR} observations is good with $\chi^2/\nu = 1.05$.

\citet{gallo+14} find that when fitting a blurred reflection model to the \textit{Suzaku} low flux spectrum that there is a degeneracy between the iron abundance and, in the model used therein, the spin parameter of the black hole (which is calculated directlty from the best-fitting inner radius of the accretion disc). They find that the spectrum can either be fit with near maximal spin ($a=0.998$) and high iron abundance (around 7 times the Solar abundance), or slightly lower spin ($a=0.94_{-0.02}^{+0.03}$ for which $r_\mathrm{in}=2.0_{-0.3}^{+0.2}$\rg) with an iron abundance only around 2 times the Solar. Including either the \textit{NuSTAR} data or simultaneously fitting all of the observed epochs, the iron abundance is constrained to the lower value, breaking this degeneracy. We find no evidence that this degeneracy exists during the other epochs, presumably because the lower energy coverage of \textit{XMM-Newton} better constrains the iron abundance as the ionisation parameter and normalisation vary. While the error bars on the accretion disc inner radius and iron abundance are larger during the \textit{Suzaku} high flux epoch, there is a single permissible range of values for each parameter. Stepping through the values of various other pairs of parameters during the fitting procedure presents no degeneracies further to those already discussed.

\subsubsection{The emissivity profile and the extent of the corona}

\begin{figure*}
\centering
\subfigure[\textit{XMM-Newton} High Flux Epoch] {
\includegraphics[width=85mm]{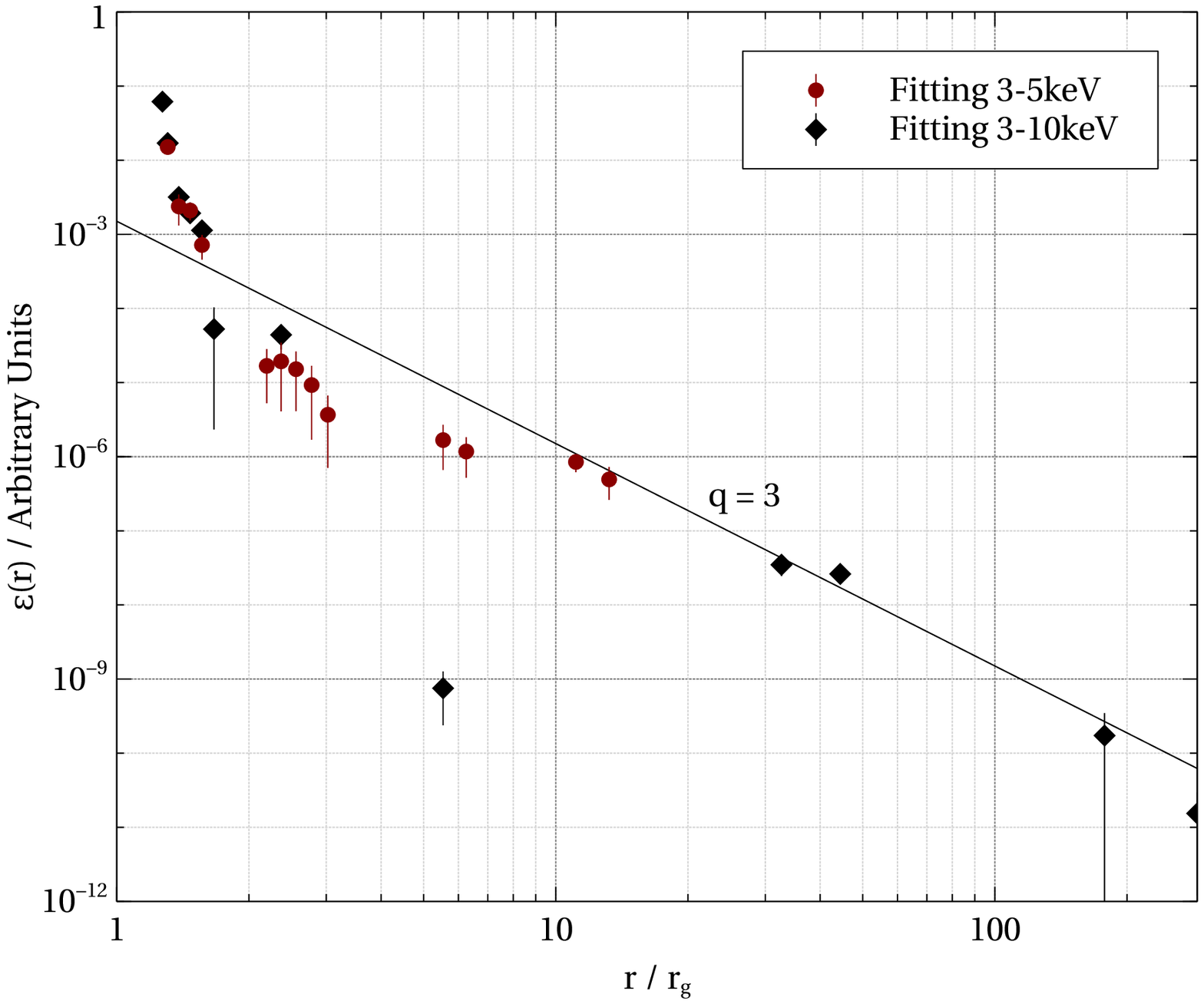}
\label{emissivity_epochs.fig:xmm_high}
}
\subfigure[\textit{Suzaku} High Flux Epoch] {
\includegraphics[width=85mm]{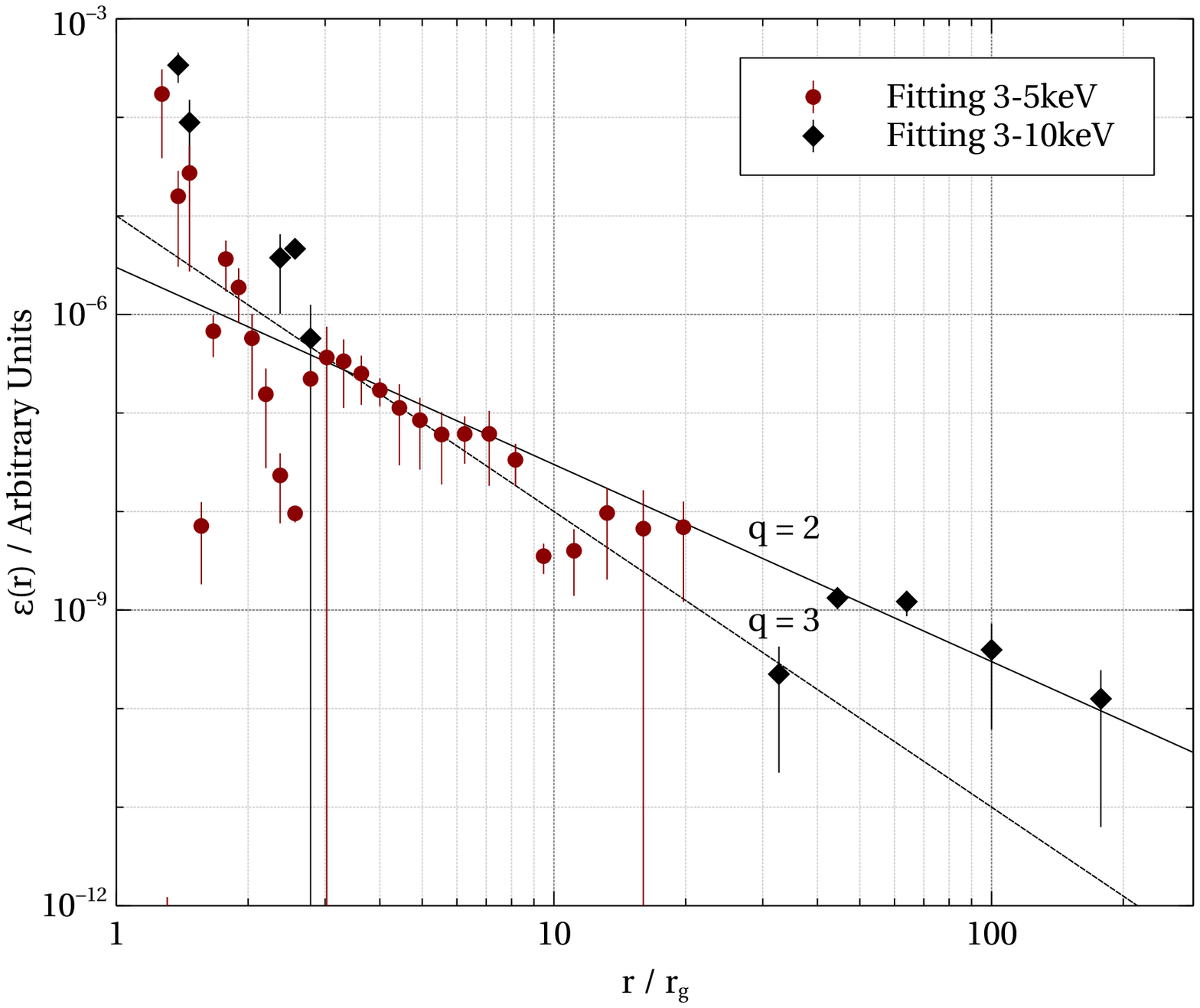}
\label{emissivity_epochs.fig:suzaku_high}
}
\subfigure[\textit{XMM-Newton} Intermediate Flux Epoch] {
\includegraphics[width=85mm]{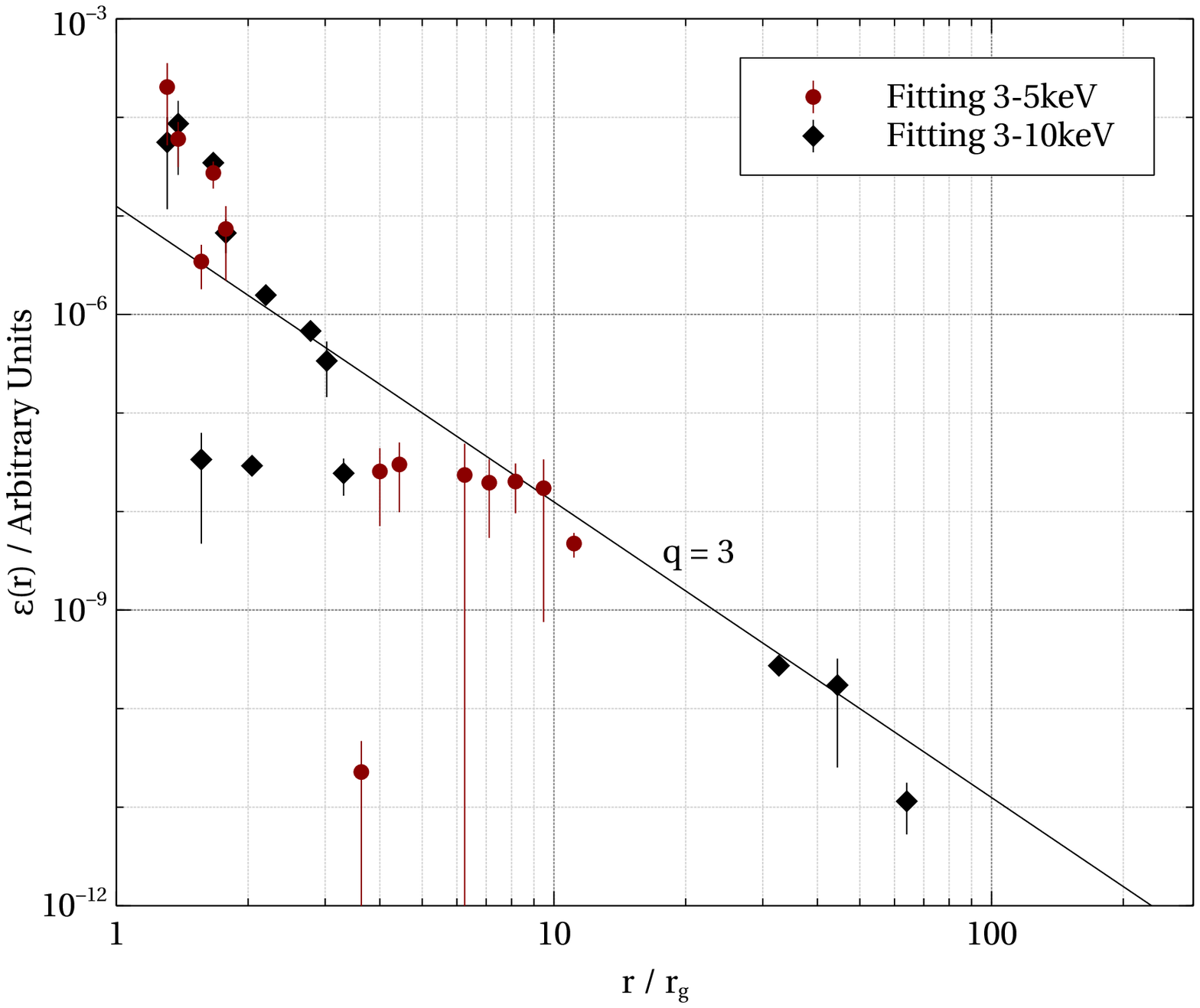}
\label{emissivity_epochs.fig:xmm_intermediate}
}
\subfigure[\textit{Suzaku} Low Flux Epoch] {
\includegraphics[width=85mm]{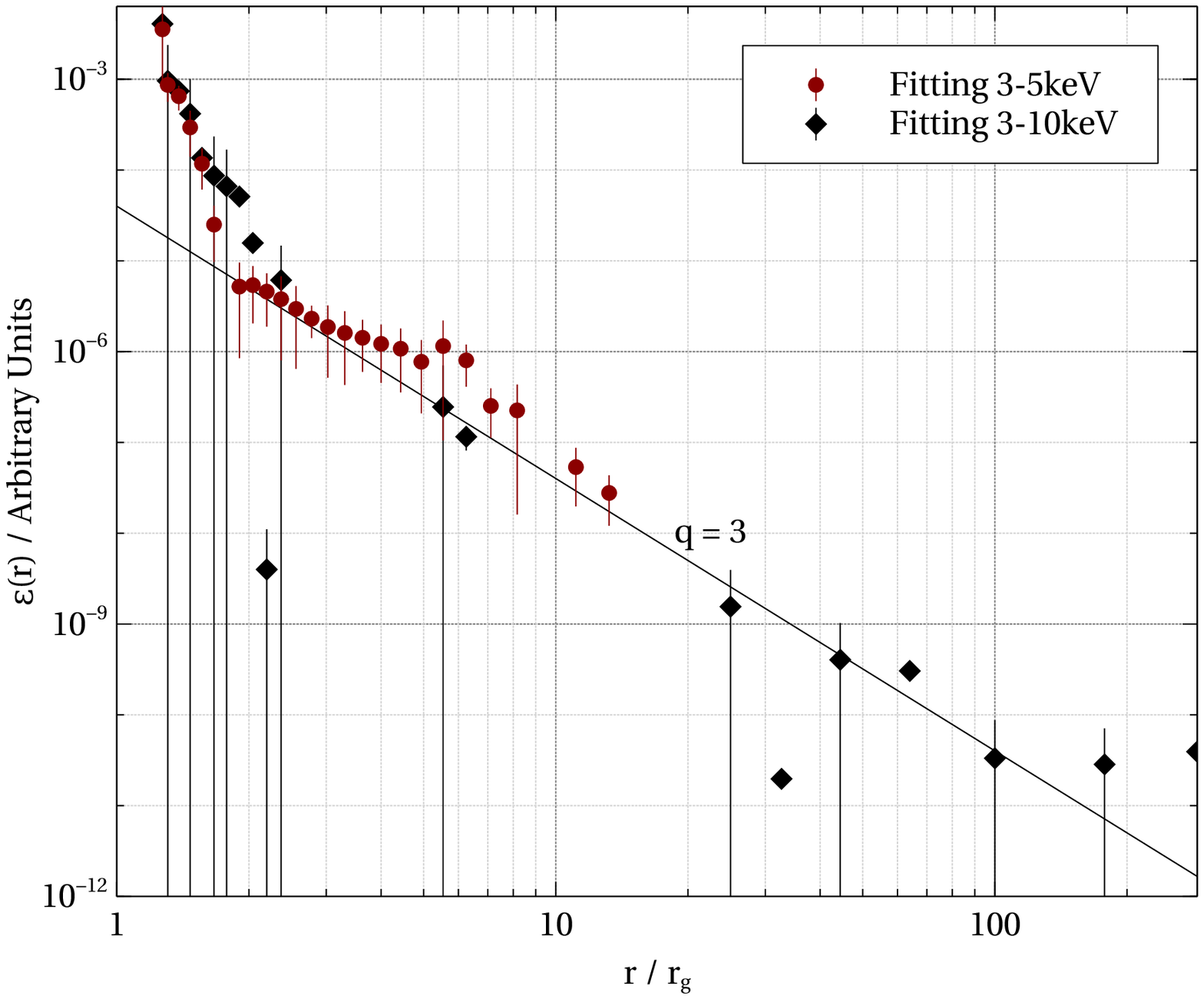}
\label{emissivity_epochs.fig:suzaku_low}
}
\caption[]{The emissivity profile of the accretion disc in Mrk~335 from the 2006 \subref{emissivity_epochs.fig:xmm_high} \textit{XMM-Newton} and \subref{emissivity_epochs.fig:suzaku_high} \textit{Suzaku} high flux, \subref{emissivity_epochs.fig:xmm_intermediate} 2009 \textit{XMM-Newton} intermediate flux and \subref{emissivity_epochs.fig:suzaku_low} 2013 \textit{Suzaku} low flux epochs. Emissivity profiles are measured by decomposing the relativistically blurred reflection from the accretion disc into the contributions from successive radii and finding the best-fitting normalisation of each component to the spectrum over both the 3-10\keV\ and 3-5\keV\ energy bands.}
\label{emissivity_epochs.fig}
\end{figure*}

Turning to the measured emissivity profiles of the accretion disc during the different epochs (Fig.~\ref{emissivity_epochs.fig}), we see great variation in the illumination of the accretion disc by the X-ray emitting corona between the high and low flux epochs. Comparing the measured profiles to the results of \citet{understanding_emis_paper} and the examples shown in Section~\ref{emissivity.sec}, we see that during the 2006 high and 2009 intermediate flux epochs observed by \textit{XMM-Newton}, the emissivity profile resembles the form expected in the case of an illuminating corona extending radially over the surface of the accretion disc at a low height; the profile approximates a twice-broken power law, falling off in this case as $r^{-9}$ over the inner part of the accretion disc, flattening at a radius of around 3\rg\ to a power law index of zero before falling off as approximately $r^{-3}$ over the outer disc. Fitting a relativistically blurred reflection component with a continuous twice-broken power law emissivity profile, using the \textsc{kdblur3} model \citep{1h0707_emis_paper}, to the high flux epoch spectrum over the energy range 1.1-10\keV\ (to include just the continuum and iron K$\alpha$ emission line) yields an outer break radius $r_\mathrm{b,out}=26_{-7}^{+10}$\rg, while for the intermediate flux epoch yields an upper limit on the outer break radius $r_\mathrm{b,out}<12$\rg\ at the 90 per cent confidence level.

There is a marked difference, however, between the 2006 high flux observations made with \textit{XMM-Newton} and \textit{Suzaku}. While the \textit{XMM-Newton} observation reveals an emissivity profile associated with a radially extended corona, flattening over the middle part of the disc before falling off as $r^{-3}$, the emissivity profile measured from the \textit{Suzaku} spectrum shows no such flattening and falls off as $r^{-2}$ (allowing up to a twice-broken power law emissivity profile, fall-off like $r^{-3}$ over the outer disc is excluded in the $6\sigma$ confidence limit). Comparing this emissivity profile to those shown in Fig.~\ref{emis_theory.fig} suggests that during this epoch, rather than being radially extended over the accretion disc, the X-ray emitting corona is extended vertically, up the rotation axis of the black hole. Theoretically, the uppermost vertical extent of the jet is marked by a subtle curvature in the outer part of the emissivity profile, from $r^{-2}$ to $r^{-3}$. The emissivity profile shows no evidence of this break within a radius of 50\rg\ suggesting significant vertical extension of the corona, however, it should be noted that as the reflected flux is falling at larger radii in the disc, this break will become harder to detect the further out it moves.

The emissivity profile measured from the 2013 observation by \textit{Suzaku} suggests that during this low flux epoch, the X-ray emitting corona is much more compact, certainly lying within 6\rg\ of the black hole, with a slight flattening of the emissivity profile observed out to this radius (although, given the error bars, this flattening is largely inferred from the two annuli at around 5.5 and 6\rg, with the other annuli within this radius consistent, within the error bars, with the $r^{-3}$ fall-off seen over the outer disc. This emissivity profile will be revisited in the next section.

We note that the emissivity profile over the outer disc is less well constrained during the lower flux epochs. The steeply-falling emissivity profile over the inner part of the disc in these cases means that very little flux is reflected from the outer part of the disc, thus we find that the emissivity profile of the outer disc in the case of the low flux epoch observed by \textit{Suzaku} is consistent with power law index $\ge 3$.

It was not possible to obtain an emissivity profile directly for the low flux epoch observed by \textit{XMM-Newton} in 2007. The low count rate combined with the short exposure time meant that there was insufficient signal-to-noise to decompose the iron K$\alpha$ emission line into the contributions from successive radii in the disc. Fitting a relativistically blurred reflection component with a twice-broken power law emissivity profile to this spectrum we find that, when the inner break radius (between the steep inner and flat parts) is frozen to 3\rg, in line with the emissivity profiles measured across the other epochs, the outer break radius is constrained to be $<4.2$\rg\ (within the 90 per cent confidence interval) and the observed spectrum is consistent with a once broken power law emissivity profile with a break radius at 3\rg\ and power law index $>2.4$ over the outer part of the disc. We therefore conclude that this low flux epoch, while not well constrained by the lower quality data, is consistent with a compact corona confined to within around 4\rg\ of the black hole.

For comparison with the findings of \citet{parker_mrk335}, the reflection spectrum was also fit using the \textsc{relxilllp} model that combines the rest-frame reflection spectrum modelled by the \textsc{xillver} code \citep{garcia+2010,garcia+2011,garcia+2013} with relativistic blurring by the \textsc{relconvlp} model that computes the relativistic blurring from an accretion disc with an emissivity profile appropriate for illumination by an isotropic point source (a `lamppost') at a variable height above the disc plane \citep{dauser+13}. In each case, except the 2006 \textit{Suzaku} high flux observation, this model was found to produce poor fits to the observed spectrum when energies below 3\keV\ are included ($\chi^2 / \nu > 1.15$) with significant residuals on the redshifted wing of the iron K$\alpha$ line in the energy range $3-5$\keV\ (it should be noted that \citealt{parker_mrk335}, who find a good fit to the 2013 \textit{NuSTAR} observation using this model, consider only energies above 3\keV, thus exclude the soft excess and the point at which the continuum meets the red wing of the iron line). The \textit{Suzaku} high flux observation is, however, well fit by this model, yielding a source height of $20_{-5}^{+7}$\rg\ above the accretion disc ($\chi^2 / \nu = 1.05$).

Using the \textsc{relxill} variant, which allows a twice-broken power law to be freely fit to the emissivity profile, rather than \textsc{relxilllp} was found to provide a good fit to the observed spectra, comparable to those using the \textsc{kdblur2} blurring kernel applied to the \textsc{reflionx} reflection spectrum. The parameters of the reflection spectrum; the iron abundance, ionisation parameter, accretion disc inclination and inner radius as well as the reflection fraction measured over the $0.1-100$\keV\ band, are found to be consistent, within the errors, between the two models.

These findings suggest that illumination of the accretion disc by a point source provides an inadequate description of the emissivity profile, highlighting the importance of measuring the emissivity profile rather than assuming a simplified model for the illuminating corona. We find that the red wing of the emission line, originating from the inner parts of the disc, is underestimated by the model assuming a point source. This suggests that the height of the point source is selected to reproduce the outer break radius of the emissivity profile, finding a high source that does not sufficiently illuminate the inner region of the disc.

\section{Short timescales: An X-ray flare during the low flux epoch}
\label{var.sec}
\begin{figure*}
\begin{minipage}{170mm}
\centering
\includegraphics[width=175mm]{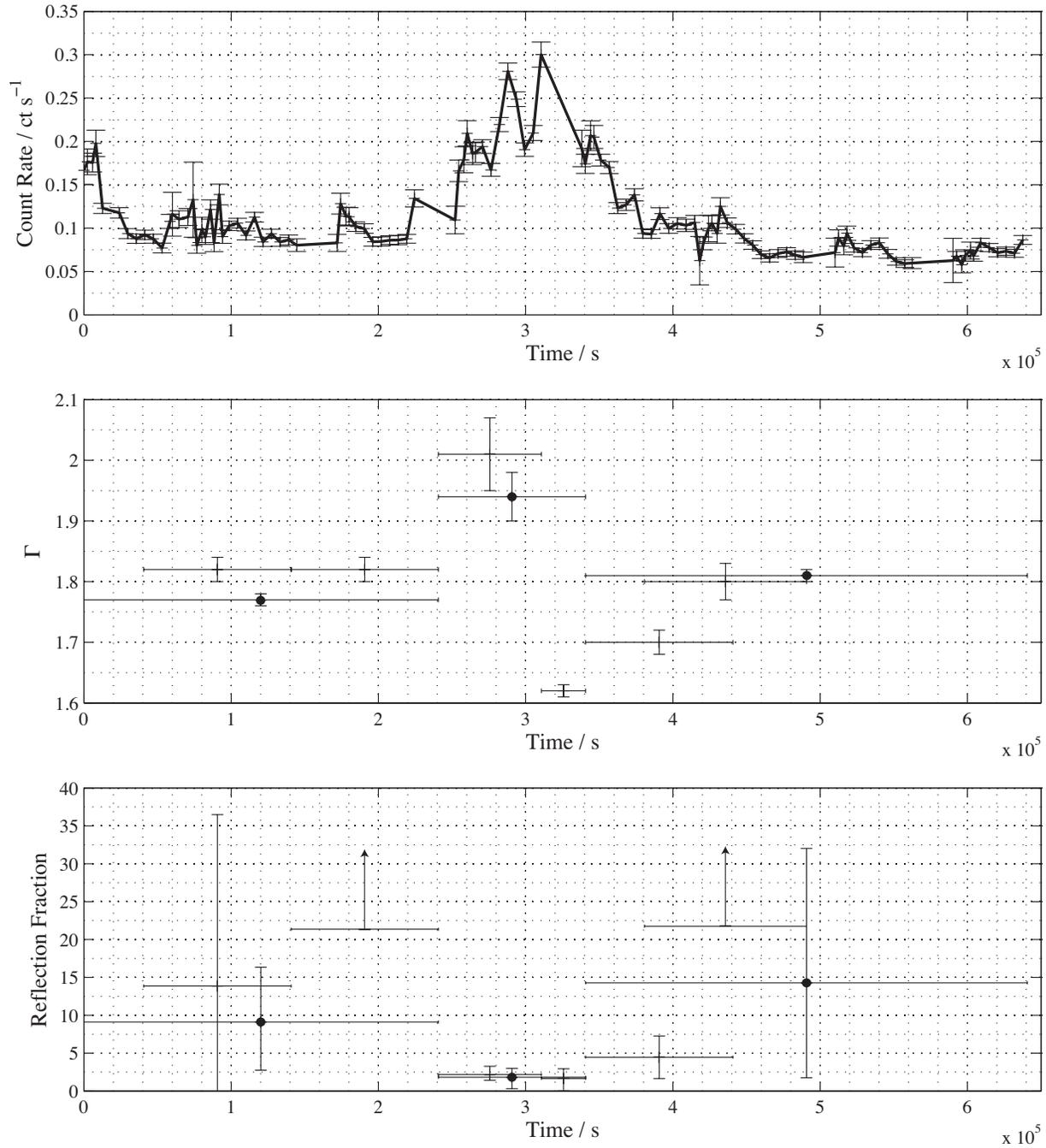}
\caption[]{Background-subtracted light curve of Markarian 335 recorded with XIS0 on board \textit{Suzaku} during the 2013 low flux epoch, showing the 90\ks\ flare approximately 250\ks\ into the observation. The light curve is averaged over each 90 minute orbit and the error bars represent the standard error in the mean. Also shown is the photon index, $\Gamma$ of the continuum spectrum and the reflection fraction, $R$, (the ratio of the photon counts in the reflection and the continuum spectral components over the 0.1 to 100\keV\ range). Points show the long time periods before, during and after the flare while crosses show the shorter subdivisions of these periods.}
\end{minipage}
\label{mrk335_flare.fig}
\end{figure*}

The light curve of the 2013 \textit{Suzaku} low state observation of Mrk~335 is shown in Fig.~\ref{mrk335_flare.fig}. Clearly apparent is a flare in the X-ray emission about 300\ks\ into the observation. The X-ray flux doubles for approximately 90\ks\ before returning to its previous level.

\subsection{Probing the short timescale variability}

In order to understand the changes in the corona that caused this rapid change in X-ray flux, the observations were divided into a number of time periods; both long periods before, during and after the flare that maximise the number of photon counts in the resulting spectra, and shorter time periods that may reveal more rapid changes.

The combined spectra measured by the front-illuminated XIS CCDs during each of the time periods were fit with the same model as the entire 2013 \textit{Suzaku} observation, with the intrinsic absorption, inclination, iron abundance and ionisation parameter frozen at the best-fit values to the spectrum from the entire 2013 low flux epoch, shown in Table~\ref{fit_results_epochs.tab}. This reduces the number of free parameters for these shorter exposures, allowing the photon index, inner radius of the accretion disc and emissivity profile, initially assumed to be a once-broken power law, using the \textsc{kdblur2} blurring kernel, to be measured for each of the time periods from the spectra constructed with fewer photon counts. 

While in reality the ionisation of the disc may vary during the flare, it can readily be shown in this model that assuming a constant ionisation parameter does not influence the measured emissivity profiles. \citet{gallo+14} measure the ionisation parameter to be less than 100\ergcmps at all times during the \textit{Suzaku} low flux observation. In this `low ionisation' regime, the spectrum is dominated by the lesser-ionised species which produce the prominent iron K$\alpha$ line at 6.4\keV. Once convolved with the relativistic blurring kernel, it is this emission line that provides the strongest measure of the emissivity profile. It is only once the ionisation parameter exceeds 150\ergcmps\ that the measurement of the emissivity profile is affected. Initially, the greater abundance of more ionised species with a vacancy in the L shell causes the iron K$\alpha$ photons to be reabsorbed, weakening the emission line and hence the statistical constraint on the emissivity profile. Finally, once helium-like and hydrogenic iron become prevalent when $\xi > 500$\ergcmps, the K$\alpha$ line is shifted from 6.4\keV\ to 6.67\keV\ and 6.97\keV, respectively and the correct emissivity profile will not be measured when assuming too low an ionisation parameter. As such, accurate determination of the ionisation state of the accretion disc only starts to become important when $\xi > 150$\ergcmps\ and becomes critical when $\xi > 500$\ergcmps (see also \citealt{1h0707_var_paper}). Moreover, since the ionised absorbing material found to be outflowing from the system imprints only narrow absorption features on the iron K$\alpha$ emission line, any short-timescale variations in this outflow will not affect the measured profile of the broad emission line, thus the assumed constancy of the absorption parameters between the time periods will not affect the measured emissivity profiles. While, of course, any variation in the ionised outflows may be of interest in understanding the changing conditions around the inner accretion flow and corona during this flare, it is difficult to constrain any variability therein during these short exposures with \textit{Suzaku}, hence we aim to reduce the number of free parameters in this work to focus on the variability in the X-ray emitting corona itself.

In order to sensitively probe the variations in the corona through changes in the photon index of the X-ray continuum, the reflection fraction and the emissivity profile of the accretion disc, after initially finding the best-fitting values of the model parameters by minimising the $\chi^2$ fit statistic in \textsc{xspec}, the probability distributions of the free model parameters during each of the time periods, given the observed data, were computed by Markov Chain Monte Carlo (MCMC) calculations as described in Section~\ref{mcmc.sec}. MCMC calculations are able to much more efficiently sample the parameter space than the conventional approach to spectral fitting, minimising $\chi^2$ and stepping through parameter values to find errors, and are much less prone to falling into local minima in the goodness-of-fit statistic, thus they are well suited to finding the best-fitting values of the model parameters to these shorter exposures from the low flux epoch and determining the range of parameter values that are admitted by the data.

A total of 24 MCMC walkers were started at the best-fitting point in the model parameter space found by minimising $\chi^2$ and run for 10,000 steps each, discarding the first 1000 steps. In order to check that the Markov chains adequately sample the parameter space, it was confirmed that the fraction of steps that were rejected due to a worsening fit was less than 75 per cent (the rule of thumb for MCMC model-fitting) and also that each of the walkers converged on the same set of parameter values by the end of their respective chains. The probability distributions for each of the free parameters were then constructed from histograms of the parameter values across the steps of all chains.

\subsection{Results}

\begin{figure*}
\centering
\subfigure[Before flare (Period A)] {
\includegraphics[width=55mm]{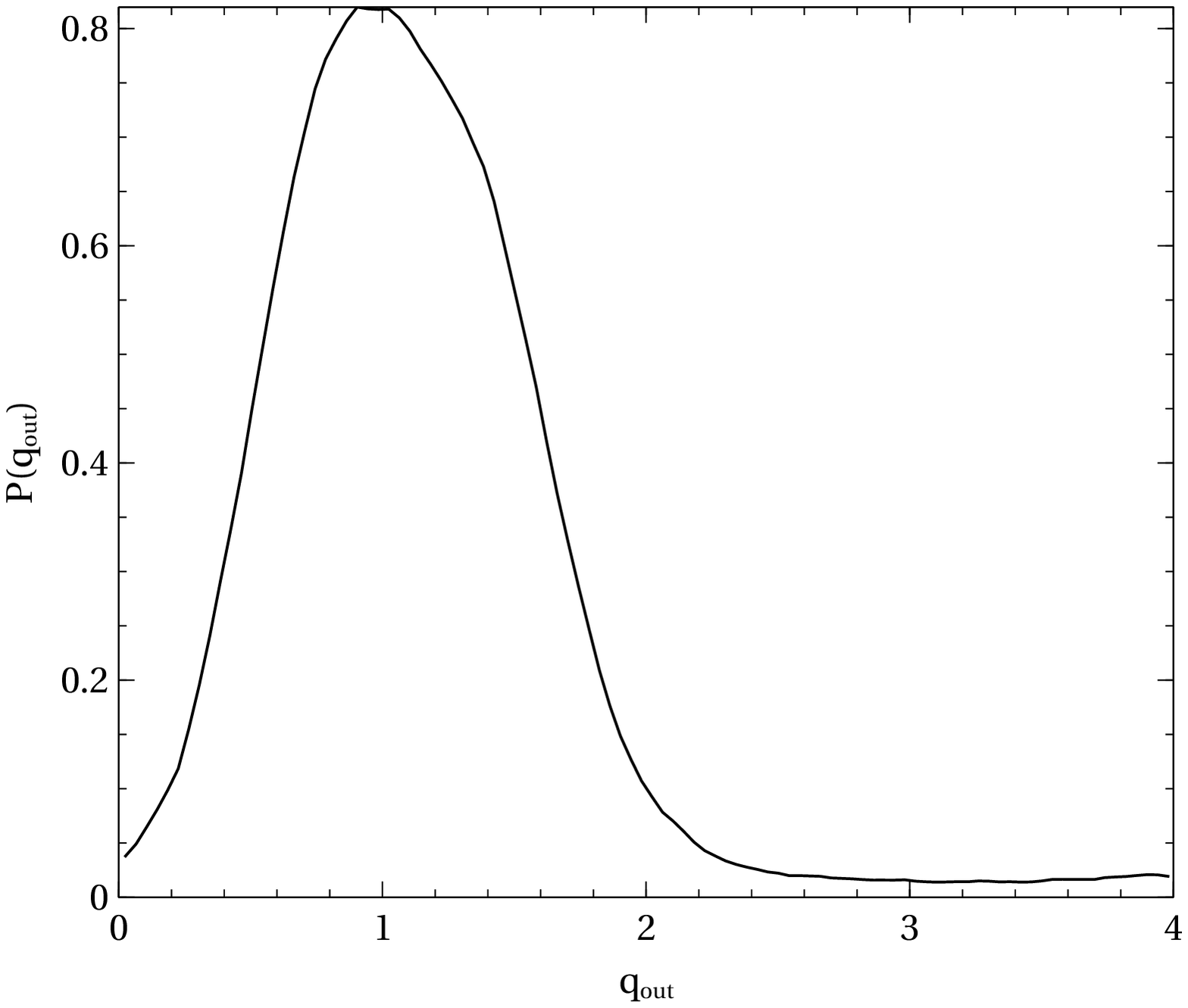}
\label{prob_qout:before_flare}
}
\subfigure[Full flare (Period D)] {
\includegraphics[width=55mm]{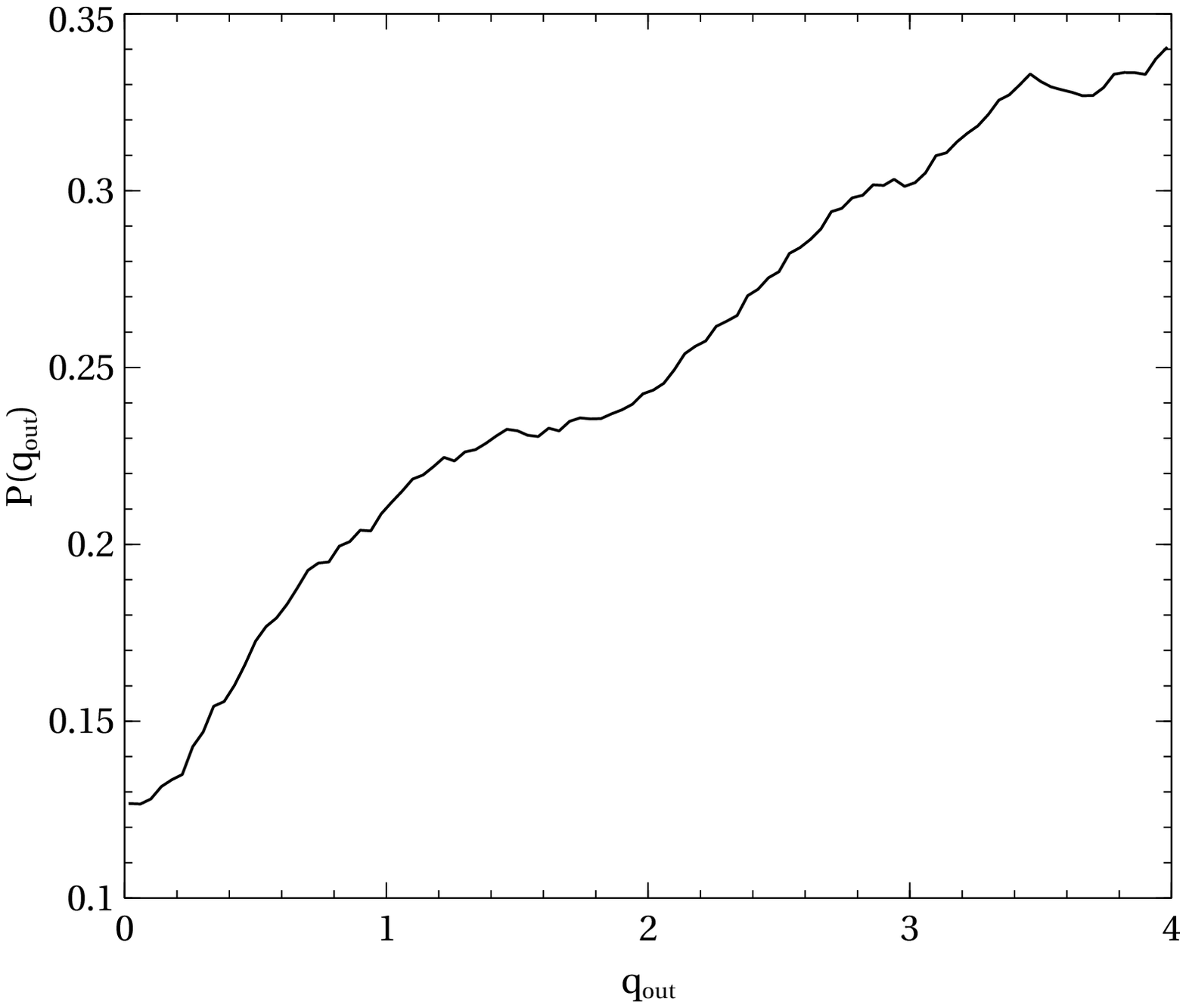}
\label{prob_qout:flare_full}
}
\subfigure[After flare (Period G)] {
\includegraphics[width=55mm]{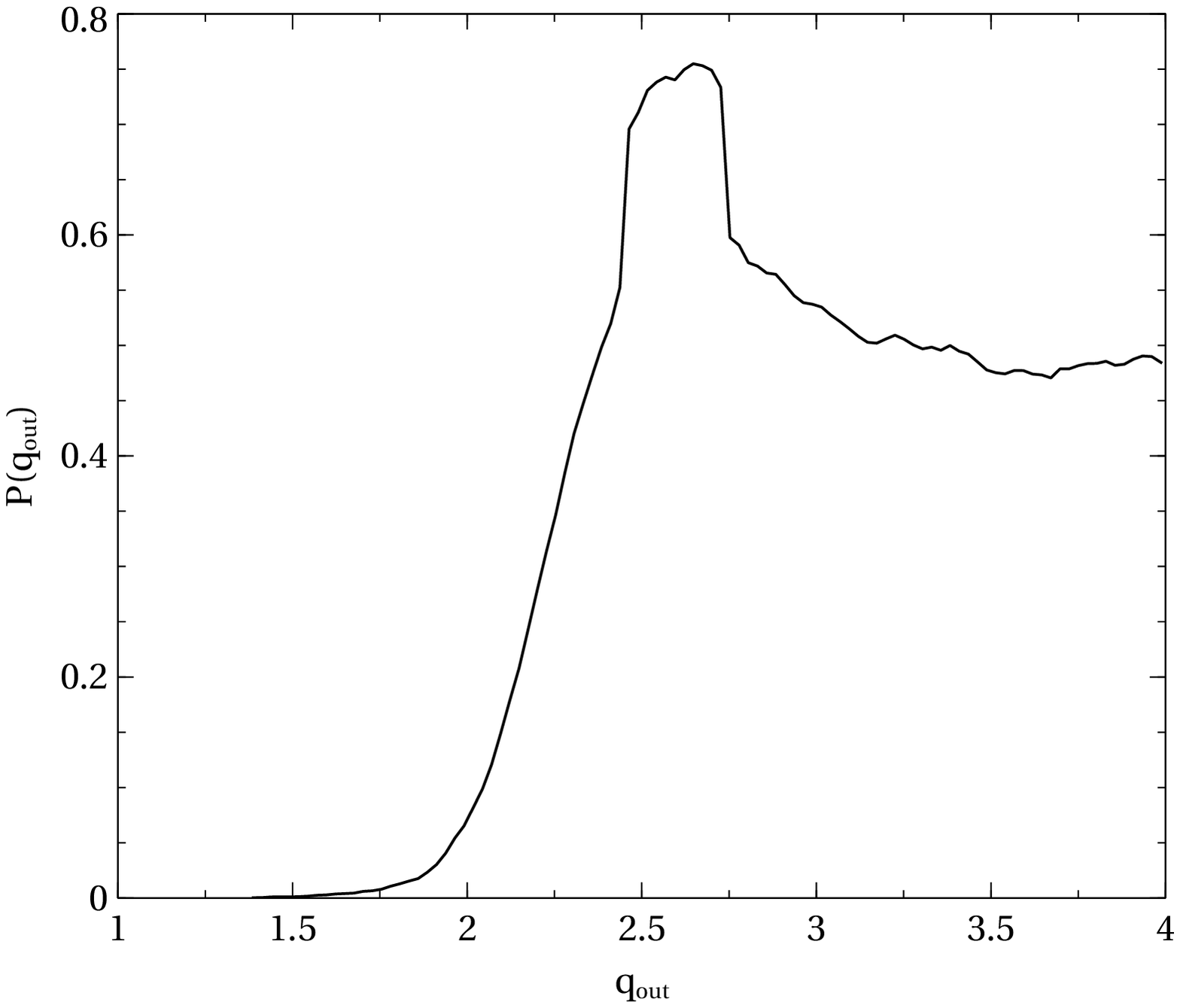}
\label{prob_qout:after_flare}
}
\caption[]{The probability distribution of $q_\mathrm{out}$, the outer slope of the once-broken power law emissivity profile of the reflection from the accretion disc given the spectra observed before, during and after the X-ray flare seen during the 2013 \textit{Suzaku} observation of Mrk~335.}
\label{prob_qout.fig}
\end{figure*}

The best fitting model parameters to the X-ray spectrum from the time periods before, during and after the flare are shown in Table~\ref{fit_results_flare.tab}, while the probability distributions of $q_\mathrm{out}$, the outer power law slope in the once-broken power law emissivity profile are shown in Fig.~\ref{prob_qout.fig}.

\begin{table*}
\begin{minipage}{140mm}
\centering
\caption{\label{fit_results_flare.tab}Best fitting values of the free parameters of the continuum and relativistically blurred reflection model components from each time period during the 2013 observation of Mrk~335 in a low flux state by \textit{Suzaku}. Parameters allowed to vary between the time periods were the reflection fraction, $R$ (defined to be the ratio of photon counts in the reflection to continuum component extrapolated over the energy range 0.1-100\keV, the photon index, $\Gamma$, of the power law continuum, the inner radius of the accretion disc and the inner power law index, $q_\mathrm{out}$ and break radius $r_\mathrm{br}$ of a once-broken power law accretion disc emissivity profile. Also allowed to vary was the outer power law index of the emissivity profile, for which probability distributions are shown in Fig.~\ref{prob_qout.fig}}
\def\arraystretch{1.5}
\begin{tabular}{lccccccc}
  	\hline
   	 \textbf{Time Period} & $R$ & $\Gamma$ & $r_\mathrm{in}$ & $q_\mathrm{in}$ & $r_\mathrm{br}$ & $\chi^2/\nu$\\
	\hline
	
	(A) Before flare & $9_{-6}^{+7}$ & $1.77\pm0.01$ & $1.26\pm0.02$ & $7.7\pm0.3$ & $7.6\pm1.5$ & 1.06 \\
	(B) Early & $14_{-13}^{+23}$ & $1.82\pm0.02$ & $1.27\pm0.02$ & $7.2\pm0.4$ & $7.0\pm1.9$  & 0.92 \\
	(C) Late & $>21$ & $1.82\pm0.02$ & $1.30\pm0.03$ & $7.8\pm0.4$ & $7.1\pm1.7$ &1.06 \\
	\hline
	(D) Full flare & $1.8_{-1.5}^{+1.2}$ & $1.94\pm0.04$ & $1.25\pm0.01$ & $9.9\pm0.1$ & $7.3\pm1.7$ & 1.01 \\
	(E) Flare rise & $2.2_{-0.7}^{+1.1}$  & $2.01\pm0.06$ & $1.25\pm0.01$ & $9.8\pm0.1$ & $7.3\pm1.7$ & 0.97 \\
	(F) Flare fall & $1.7_{-1.6}^{+1.3}$ & $1.62\pm0.01$ & $1.29\pm0.03$ & $9.5\pm0.4$ & $6.8\pm2.0$ & 1.43 \\
	\hline
	(G) After flare & $14_{-12}^{+18}$ & $1.81\pm0.01$ & $1.25\pm0.01$ & $8.5\pm0.3$ & $3.4\pm0.3$ & 1.03 \\
	(H) Immediately after 1 & $4_{-3}^{+3}$   & $1.70\pm0.02$ & $1.27\pm0.02$ & $9.3\pm0.4$ & $4.3\pm1.7$ & 1.12 \\
	(I) Immediately after 2 & $>22$  & $1.80\pm0.03$ & $1.27\pm0.02$ & $9.1\pm0.4$ & $3.7\pm1.0$ & 0.95 \\
	\hline
\end{tabular}
\end{minipage}
\end{table*}

There is a clear change in the geometry and energetics of the corona during the X-ray flare, indicated by a significant decrease in the reflection fraction and increase in the photon index of the continuum during the flare before they return to their previous values. This behaviour is consistent with the long timescale variability; a softening continuum spectrum and decreased reflection fraction, suggesting a cooling and expanding corona as the X-ray flux increases. The measured inner radii of the accretion disc from all time periods before, during and after the flare show no evidence for movement of the inner edge of the disc or truncation thereof.

Looking at the best-fitting once-broken power law emissivity profile to each of the spectra, there is a clear change in the corona between the time periods before and after the flare. In all cases, the emissivity profile is found to fall steeply over the inner part of the accretion disc, with the power law index tightly constrained to being greater than 7 in each case. The break radius is found to be around 7\rg\ during each time period before and during the flare, but then decreases to around 4\rg\ after the flare. The most notable difference, however, is in the observed slope in the emissivity profile of the outer part of the accretion disc, which, from Section~\ref{emis_fit.sec}, is the most sensitive to the spatial extent of the corona.

The probability distributions of this parameter in Fig.~\ref{prob_qout.fig} show that before the flare, the outer section of the emissivity profile is significantly flatter than $q_\mathrm{out}=3$, which we see from Section~\ref{emissivity.sec} would be expected not only in the classical case, but in the case of a compact X-ray source confined close to the black hole. The measurement of a power law index $<2$ over the outer part of the disc is suggestive of flattening of the middle section of the emissivity profile and combining this with the observation that the inner part of the profile is steep suggests that the corona during this time period is spatially extended.

\begin{figure*}
\centering
\subfigure[Before Flare] {
\includegraphics[width=85mm]{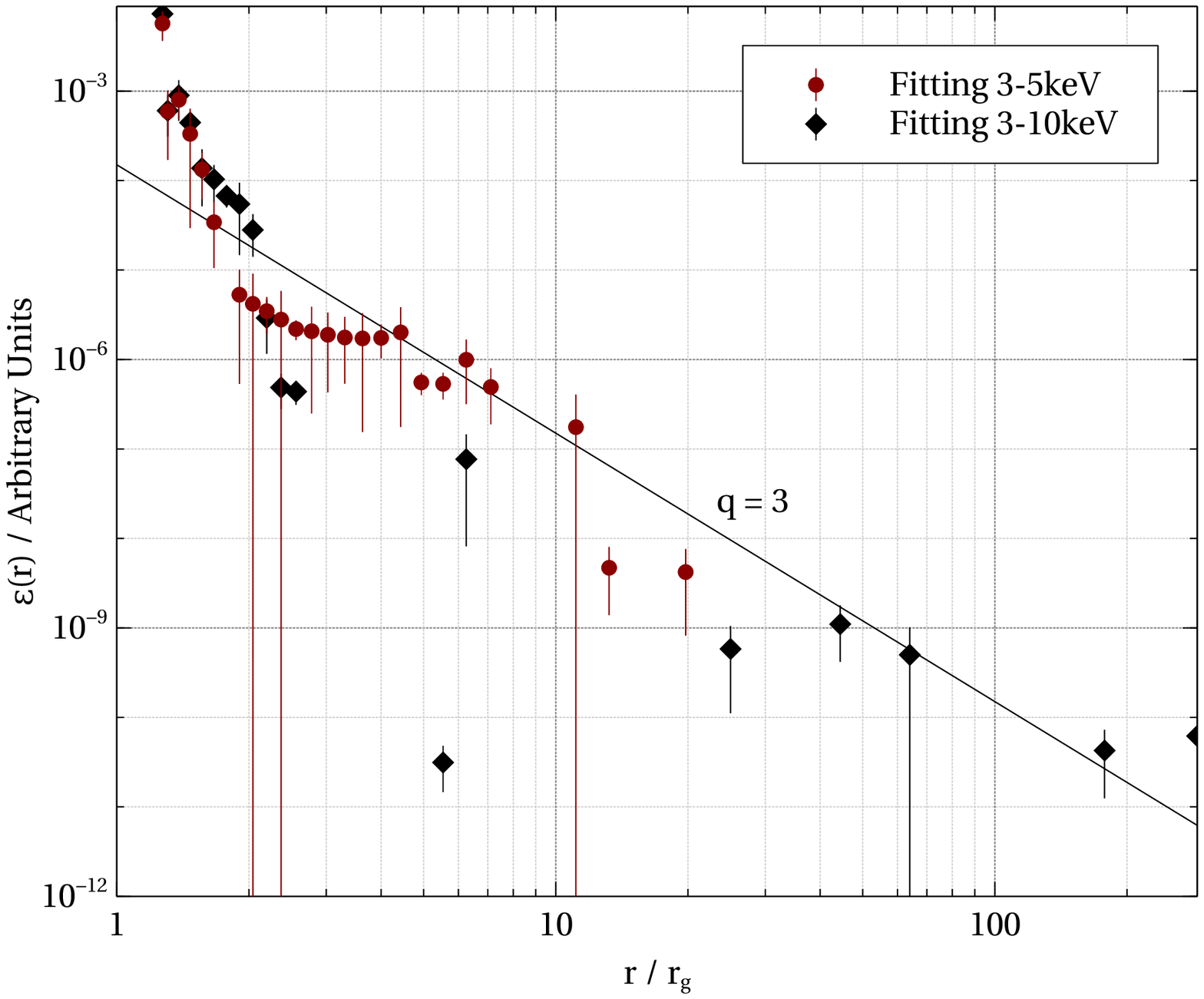}
\label{emis_flare.fig:before}
}
\subfigure[After Flare] {
\includegraphics[width=85mm]{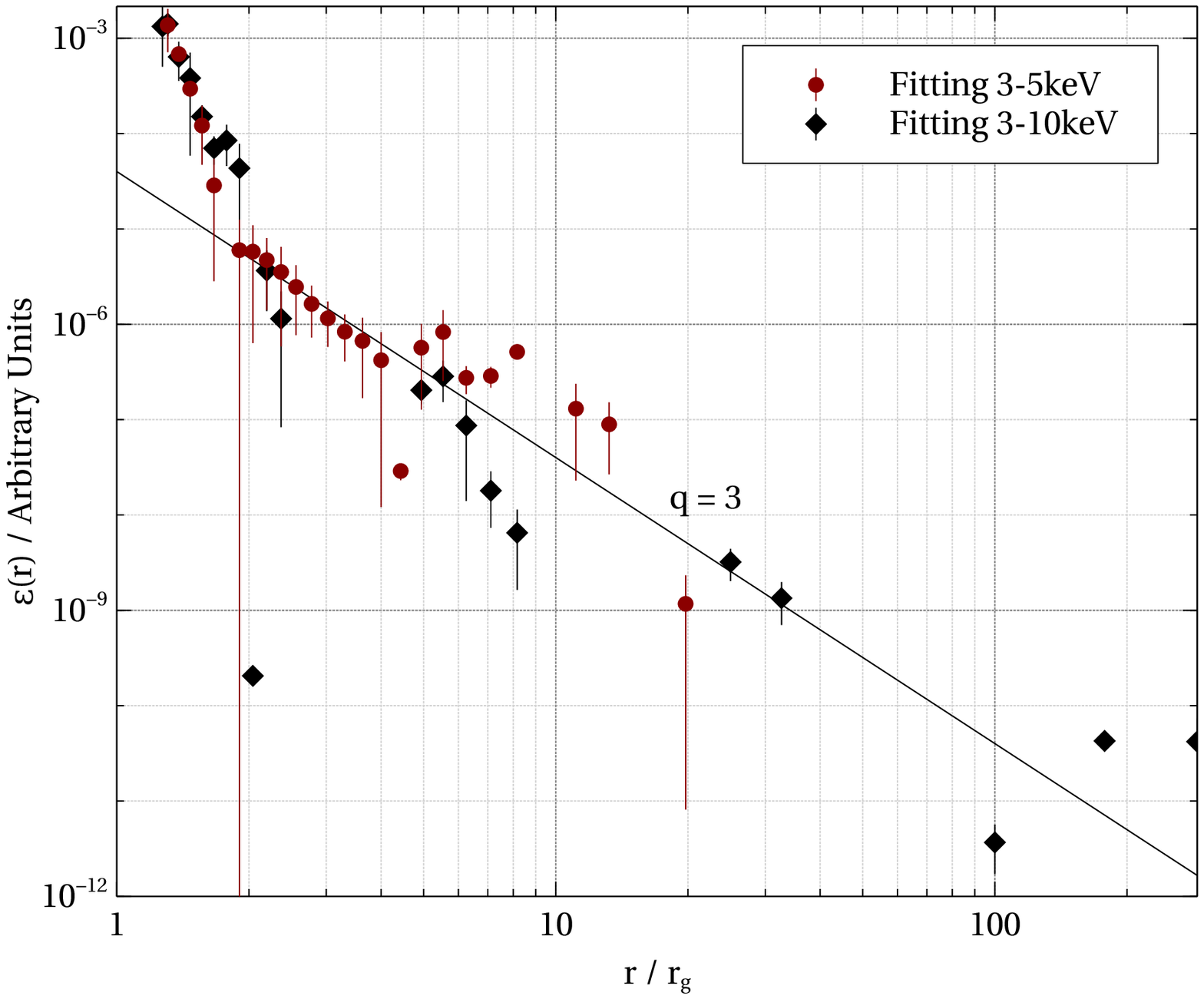}
\label{emis_flare.fig:after}
}
\caption[]{The emissivity profile of the accretion disc from the periods \subref{emis_flare.fig:before} before and \subref{emis_flare.fig:after} after the flare in the 2013 low flux epoch observed by \textit{Suzaku} measured by decomposing the relativistically blurred reflection from the accretion disc into the contributions from successive radii over both the 3-10\keV\ and 3-5\keV\ energy bands.}
\label{emis_flare.fig}
\end{figure*}

After the flare, we clearly see that the power law index of the emissivity profile over the outer disc becomes significantly steeper, shifting from $q_\mathrm{out}\lesssim 2$ before the flare to $q_\mathrm{out}\gtrsim 2.5$ afterwards. In order to better understand this change in the best-fitting power law indices, the emissivity profiles of the accretion disc were measured directly from the profile of the iron K$\alpha$ emission line using the method of \citet{1h0707_emis_paper}, detailed in Section~\ref{measure_emis.sec}, from the total combined spectra before and after the flare. The measured emissivity profiles are shown in Fig.~\ref{emis_flare.fig}. We see that the apparent flattening seen in the average emissivity profile from the whole \textit{Suzaku} low flux epoch observation in Fig.~\ref{emissivity_epochs.fig} results entirely from the period before the flare. The flattening seen in the emissivity profile before the flare is consistent with a corona extending over the inner regions of the accretion disc to a radius no further than 5\rg, while the high reflection fraction implies it extends only a few gravitational radii vertically, above the disc. On other other hand, the emissivity profile after the flare consists of only a steep inner part, transitioning to a fall-off slightly steeper than $r^{-3}$ with no flattening, indicating a compact corona within just 2\rg\ of the central black hole. This conclusion is further supported by the extremely high reflection fraction ($R>22$) measured after the flare. Thus, we see that the X-ray flare marks a reconfiguration of the corona from an extended entity, covering the inner part of the accretion disc, to a much more compact source of X-ray emission.

During the flare itself, the reflection fraction drops significantly with the extra X-ray flux during the flare largely seen directly in the continuum, rather than in the reflection from the accretion disc. The reflected photon count exceeds that seen in the continuum by a factor of more than 10 before and after the flare, while during the flare itself, the enhancement in the continuum reduces the reflection fraction to approximately 2.

It is difficult to constrain the outer power law index of the emissivity profile during the flare since the flare is short and the reflection fraction is diminished. A steeper index is preferred by the data during the whole period of the flare (pegging at the upper limit of 4), however this distribution is broad and the outer power law index is steepened only within $1\sigma$ from that during the pre-flare period.

The significant drop in the reflection fraction, combined with the apparent steepening emissivity profile is consistent with the corona becoming collimated, during the flare, into a jet-like configuration, while mildly relativistic outward motion of material in the corona in this jet-like configuration can help to explain such a dramatic drop the reflection fraction from such a compact corona \citep[\textit{e.g.}][]{beloborodov}. Since the corona is now covering less of the inner accretion disc, we expect the emissivity profile to steepen to a power law index between around 2 and 3 over the outer part of the disc (where the break from 2 to 3, or, in this case, the average value measured, depends upon the maximum vertical extent of this collimated, jet-like corona).

We note a poorer fit to  the spectrum during time period F, the decline of the flare. This is due to excess emission (at approximately the $1.5\sigma$ level) in four spectral bins between 3.0 and 3.1\keV\ that appears only during this period. This time period is too short, however, for there to be sufficient photon counts to determine the cause of this excess.

\begin{figure*}
\begin{minipage}{170mm}
\centering
\includegraphics[width=175mm]{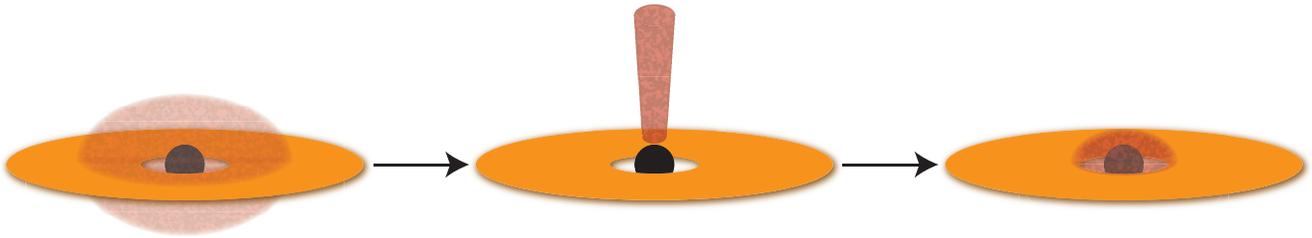}
\caption[]{The inferred evolution of the geometry of the X-ray emitting corona in Mrk~335 during the flare during the 2013 low flux epoch observation with \textit{Suzaku}. The emissivity profile of the accretion disc before and after the flare show that the event corresponds to a reconfiguration of the corona from a slightly extended to compact system, confined within around 2\rg\ of the black hole, while the sharp drop in the reflection fraction and the probability distribution of the best-fitting power law indices to the emissivity profile suggest that the flare itself is linked to a vertical collimation of the corona into a jet-like configuration before dissipating.}
\end{minipage}
\label{mrk335_flare_evolution.fig}
\end{figure*}

During the flare, the continuum spectrum softens, with the photon index  increasing from $\Gamma=1.77\pm 0.01$ before the flare to $2.01\pm 0.06$ during the flare, while dividing the observation into shorter time periods during and immediately after the flare shows that during the decline of the flare, the continuum spectrum actually hardens to $\Gamma=1.62\pm 0.01$, harder than the average photon index measured over the period following the flare. Combining this observation with the increases in the reflection fraction immediately before (Period C) and after the flare (Period I) suggests that the corona is compressing before it outbursts into the flare when it cools, softening the continuum spectrum. Then, as the flare declines, the corona becomes extremely compact and hot, increasing the reflection fraction and decreasing the photon index. Finally, it expands and cools again slightly as the spectrum softens slightly to the pre-flare photon index. The evolution of the corona during the flare is illustrated in Fig.~\ref{mrk335_flare_evolution.fig}.

\section{Discussion}
Looking at the variability of Mrk~335 on long timescales, we observe two major patterns in the evolution of the X-ray emitting corona as the source transitions from lower to higher flux epochs. In general, as the X-ray luminosity increases, the corona expands and the X-ray continuum it emits softens. The corona was found to expand from a compact configuration, confined to within just $2\sim 3$\rg\ of the black hole, during the lowest flux epochs, becoming more extended (to less than 12\rg\ around the black hole) in the intermediate flux epoch seen in 2009, then expanding to a much more extended configuration during high flux epochs, reaching 26\rg\ over the accretion disc during the 2006 \textit{XMM-Newton} observation.

During this observation, \citet{kara+13} measure a time lag of $\sim 150$\s\ between correlated variations in energy bands dominated by the primary X-ray continuum and by its reflection from the accretion disc. Interpreting this as the light travel time between the primary X-ray source and the reflecting accretion disc and taking the mass of the black hole to be $2.6\times 10^7$\Msun\ \citep{grier+12} allows the average height of the corona vertically above the accretion disc to be calculated as $\sim 1.5$\rg\ \citep{lag_spectra_paper}.

Ray tracing simulations show that for such a radially extended corona, the reflection fraction is insensitive to the vertical extent of the corona (so long as it lies below 10\rg) so the corona was taken to extend from 1\rg\ to 2\rg\ above the disc to be consistent with the lag. For a corona extending to 19\rg, the lower bound of the measured extent at the 90 per cent confidence level, the ratio of the number photons from the corona that are incident upon the disc to that which is able to escape is 1.45. When photons are emitted from an accretion disc with the measured emissivity profile, only 72 per cent of those photons were found to be able to escape to be detected in the reflection spectrum (including those that are returned to the disc to give higher order reflections), thus predicting a reflection fraction of 1.04, slightly underestimating the measured value of $1.3_{-0.2}^{+0.4}$. We find, however, that if the central regions of the corona are more luminous than the extremities, as might be expected if more energy is liberated from the inner regions of the accretion flow, even with just a slight luminosity gradient, proportional to $r^{-0.5}$, the observed reflection fraction is consistent with the inferred extent of the corona and no such a steep gradient as to affect the measured emissivity profile, though a more complete study of the reflection fraction expected from extended coron\ae\ will be conducted in future work.

During the intermediate flux epoch, the emissivity profile constrained the radial extent of the corona to being less than 12\rg\ over the surface of the disc, however the measured reflection fraction of $1.8_{-0.3}^{_0.3}$ further constrains the radial extent of the corona to being within 6\rg\ if the luminosity is constant throughout the corona. The radial extent of the corona is still constrained only to within 12\rg, however, if the corona is brighter in the central parts.

As the corona expands into the higher flux epochs, the continuum spectrum softens with an increasing photon index (except for the anomalous value measured during the 2007 low flux epoch observed by \textit{XMM-Newton}). The increasing X-ray flux from an expanding corona suggests that the optical depth experienced by seed photons from the accretion disc is not decreasing, therefore the softening in the continuum spectrum, from $\Gamma=1.9$ to $\Gamma=2.5$, following \citet{1h0707_var_paper}, is caused by an increase in the temperature of the corona from the high to low flux epochs of around 90 per cent, assuming the optical depth remains constant.

While the spectra appear similar by eye, we find that the two high flux epochs measured by \textit{XMM-Newton} and \textit{Suzaku} in 2006, correspond to very different configurations of the corona. During the \textit{XMM-Newton} observation, the corona was found to be radially extended over the inner part of the accretion disc, spanning to a radius of around 26\rg. On the other hand, during the \textit{Suzaku} observation later that year, the observed spectrum suggests that the corona has become vertically collimated into a jet-like configuration up the rotation axis of the black hole, perpendicular to the disc plane. During this observation, the reflection fraction (the ratio of photons reflected from the accretion disc to those detected in the continuum) was found to be 0.27. This is significantly less than unity, as would be expected in the case of either a greatly extended source or point source located at a point a large distance from the black hole where an observer would see the accretion disc subtending a solid angle of $2\pi$, thus half of the emitted rays should be reflected and half observed directly. Seeing such a low reflection fraction from a vertically collimated, jet-like corona suggests that material in the corona may be moving relativistic speeds up the jet-like structure. This would cause the continuum emission from the corona to be beamed into the direction of motion, away from the disc, explaining the low reflection fraction measured. This vertically collimated phase of the corona is not long-lived, with the source dropping back into a low flux state within just a year, before it was observed by \textit{XMM-Newton} in July 2007.

\citet{dauser+13} (see also \citealt{understanding_emis_paper}) show, however, that relativistic motion of the jet causes a reduction in the emissivity of the inner part of the accretion disc as the beaming of emission upwards in the jet leads to less illumination of this region. This can cause emission lines from the disc to appear narrower and give the appearance of a truncated accretion disc. In this instance, we clearly detect reflected emission from the innermost regions of the accretion disc and, moreover, measure a steeply falling emissivity profile over the inner parts. To reconcile these results, we infer that the base of the jet-like structure is not moving relativistically such that it can illuminate the inner regions of the accretion disc, and that the material is accelerated as it travels up the lower part of the jet. \citet{gallo+14}, following \citet{beloborodov}, show that the measured reflection fraction is consistent with a jet velocity $\beta = v/c = 0.28$. This is the escape velocity at 25\rg\ from the black hole. Mrk~335 is generally considered to be radio-quiet (although there were no radio observations during this epoch), suggesting that the material does not escape from the corona and is not able to form extended jets and lobes characteristic of radio galaxies \citet{ghisellini+04}. We therefore conclude that the base of the corona should be slow-moving and, if we are to assume that there was no radio emission during this epoch, the material has been accelerated to $\beta = 0.28$ within 25\rg\ of the black hole. The material is able to continue moving up the jet, since the emissivity profile of the accretion disc suggests illumination by a corona extending at least 50\rg\ from the black hole, though presumably in the upper parts of the jet, the material is decelerated and brought back down by the gravitational attraction of the black hole.

Intrinsic absorption by outflowing material, possibly a wind arising from the accretion disc, is detected at some level during all epochs \citep{longinotti+13}, except the 2006 high flux epoch observed by \textit{Suzaku} \citep{larsson+08}. We note that it is when the corona is found to have become collimated into a jet-like structure that the outflow is lost. \citet{king_wind_jet} report an apparent dichotomy between black holes across the mass scale between dominance of the jet power and the power in outflowing winds, with a transition occurring at $L_\mathrm{Bol}\sim 10^{-2} L_\mathrm{Edd}$ with the jet power dominating at lower luminosities and the wind power dominating at higher luminosities. They also report that jets and winds obey the same scaling relations to the luminosity, suggesting they may be driven by the same underlying process. It is possible that we have observed Mrk~335 undergo this transition and a change in the mass accretion rate (leading to a change in the bolometric luminosity) that accompanied the collimation of the corona and the loss of the outflow.

Underlying this general behaviour on long timescales is more complex variability, exemplified by an X-ray flare observed during the low flux epoch observed by \textit{Suzaku} in 2013. During this flare the X-ray flux doubles for a period of 90\ks\ before returning to its previous level. Measurements of the emissivity profile of the accretion disc before and after this event show that it marks a reconfiguration of the X-ray source from an extended corona (albeit quite small, spanning to a radius of only 5\rg) to a much more compact source within just $2\sim 3$\rg\ from the black hole.

While the emissivity profile of the accretion disc during the flare itself is not well constrained, the flare being quite short in duration, the probability distribution parameters of the broken power law emissivity profile suggest an initial steepening of the emissivity profile as the flare begins, prior to the final steepening as the corona collapses down to the compact configuration following the flare. The steepening of the emissivity profile into the flare, combined with the dramatic reduction in the reflection fraction is suggests that the corona becomes collimated into a jet-like configuration for a short time before dissipating.

The average reflection fraction measured during the 2013 low flux epoch suggests that, on average, the corona was confined to within 1.5\rg\ of the black hole, while in order to explain the extremely high reflection fractions seen immediately before and after the flare, it would appear that with corona extended just 0.5\rg\ radially from the rotation axis of the black hole, while being within just 0.3\rg\ vertically of the event horizon. During the flare itself, the reflection fraction reducing to $1.8_{-1.5}^{+1.2}$ suggests that the corona extended vertically more than 4\rg\ from the black hole, though given the uncertainty in the measurement, it is not possible to constrain the upper bound of the extent of the corona.

While it is generally believed that jets are launched from the accretion flows onto black holes by magnetic fields anchored to the disc that become twisted into a jet-like structure, accelerating particles along their length \citep{blandford_rees}, the exact mechanism by which energy is released and the jets are accelerated remains unknown. The model of \citet{blandford_znajek} suggests that magnetic field lines threading the event horizon of the black hole are able to tap energy from the spin of the black hole to drive the jets while \citet{blandford_payne} suggest that the jets are launched by the magneto-centrifugal acceleration of particles by the magnetic field lines anchored to the orbiting disc material. \citet{mckinney+2012} studied a series of general relativistic magneto-hydrodynamical simulations. They find that toroidal magnetic fields are accreted inwards to form a strongly magnetised, transient dipolar jets, while poloidal magnetic fields are readily accreted and cause the inner part of the disc to become compressed, before a critical magnetic flux density is reached and the magnetic field undergoes an inversion and the field responsible for compressing the inner part of the disc is lost.

\citet{gallo+14} demonstrate that an observed correlation between hardness ratio and count rate that appears only in the time period following the flare can be explained by an increase in the ionisation of the accretion disc as the X-ray count rate subsides. This can be explained by the compression of the disc by the accreted poloidal magnetic field as the flare begins and the corona begins to launch into the jet-like structure and then the loss of the magnetic field following the inversion, allowing the compressed disc to expand again. The decreasing density of the inner accretion disc reduces the recombination of electrons and ions so, subject to illumination by similar levels of X-ray flux from the corona, the disc reionises. During the time periods immediately following the flare, we notice that the reflection fraction becomes extremely high and the continuum spectrum hardens to a lower photon index than is observed either before the flare or the average of the whole period after the flare, suggesting that immediately following the fall of the flare, the X-ray has collapsed into a small but hot corona, before cooling (although remaining compact) toward the end of the observation.

\section{Conclusions}
The long and short timescale variability in the X-ray emitting corona of Markarian 335 was studied through measurement of the emissivity profile of the accretion disc, the varying fraction of X-rays emitted from the corona that are reflected from the accretion disc and the photon index of the X-ray continuum.

On long timescales, the X-ray luminosity is seen to vary by more than an order of magnitude, corresponding to an expansion of the corona to a fill a larger volume, covering the inner regions of the accretion disc out to around 26\rg\ during the highest flux epoch observed in 2006, then contracting to within 12\rg\ and 5\rg\, respectively, during 2009 intermediate and 2013 low flux epochs. This expansion corresponds to a cooling of the corona, with a softening of the continuum spectrum from $\Gamma=1.9$ during the the 2013 low flux epoch to $\Gamma=2.5$ in the 2006 high flux epoch.

While the earlier 2006 high flux epoch, observed by \textit{XMM-Newton}, is well described by a corona extending to a low height, over the surface of the accretion disc, the measured emissivity profile of the later 2006 high flux epoch observed by \textit{Suzaku} suggests a vertically-collimated jet-like corona with the low measured reflection fraction suggesting relativistic motion of the X-ray emitting material away from the black hole.

Underlying this time-averaged behaviour on long timescales are more complex behaviours on short timescales. Measurement of the emissivity profile of the accretion disc before an X-ray flare during the 2013 low flux observation shows that this flare corresponded to a reconfiguration of the X-ray source from a slightly extended corona reaching around 5\rg\ over the surface of the disc before the flare to a much more compact configuration within approximately $2\sim 3$\rg\ after the flare. There is no evidence for truncation of the accretion disc during the flare, with the inner radius of the accretion disc consistent with the innermost stable circular orbit (ISCO) of a maximally spinning black hole at all times during the observation.

The variation in the reflection fraction as well as a steepening in the best fitting index of a power law fit to the emissivity profile during the flare may suggest that the flare itself corresponds to an `aborted jet launch' during which the corona becomes collimated and vertically extended for a short time before collapsing into the compact post-flare configuration, with the continuum spectrum hardening beyond the pre-flare spectrum, before finally returning to the original photon index.

Low flux epochs in AGN represent unique opportunities to study the environments in the strong gravitational fields close to black holes, allowing for clear observation of relativistically blurred reflection from the inner few gravitational radii on the accretion disc, with less dilution by the outer disc and X-ray continuum. Observing the changes in the physical properties of the X-ray emitting corona; its geometry and energetics, will enable comparisons to be made between observed AGN and theoretical predictions of coron\ae\ arising close to supermassive black holes from general relativistic magneto-hydrodynamical (GRMHD) simulations and will shed further light on the physical processes that inject the gravitational energy from the accretion flow into X-ray emitting coron\ae\ to power some of the brightest persistent sources of electromagnetic radiation in the Universe.

\section*{Acknowledgements}
DRW is supported by a CITA National Fellowship. This work is based on observations obtained with \textit{XMM-Newton}, an ESA science mission with instruments and contributions directly funded by ESA Member States and NASA and has also made use of data obtained from the \textit{Suzaku} satellite, a collaborative mission between the space agencies of Japan (JAXA) and the USA (NASA). This work also uses observations made by the NuSTAR mission, a project led by the California Institute of Astronomy, managed by the Jet Propulsion Laboratory, and funded by NASA. We thank the anonymous referee for their constructive feedback on the original manuscript.

\vspace{-0.4cm}
\bibliographystyle{mnras}
\bibliography{agn}

\label{lastpage}

\end{document}